\documentclass[12pt]{JHEP3}
\usepackage{nicefrac}
\usepackage{graphicx}
\usepackage{subfigure}
\usepackage{epsfig}
\title{Intersecting Flavor Branes}

\preprint{ YITP-SB-09-45}

\author{Elli Pomoni\footnote{Email: elli.pomoni@stonybrook.edu } $\,$  and Leonardo Rastelli\footnote{Email: leonardo.rastelli@stonybrook.edu}
\\ \\ \\
\it  C.N. Yang Institute for Theoretical Physics,\\
\it Stony Brook University, \\
\it Stony Brook, NY 11794-3840, USA}

\abstract{ 
\smallskip

We consider an instance of the AdS/CFT duality where the bulk theory contains an open string tachyon, and study
the instability from the viewpoint of the boundary field theory. 
We focus on the specific example of the  $AdS_5 \times S^5$ background with two probe  $D7$ branes intersecting at general angles.
 For generic angles supersymmetry is completely broken and there is an open string tachyon between
 the branes. The field theory action for this system is obtained by coupling to ${\cal N} =4$ super Yang-Mills
 two ${\cal N} =2$ hyper multiplets  in the fundamental representation of the $SU(N)$ gauge group, but with  different
choices of embedding of the two ${\cal N}=2$ subalgebras into ${\cal N}=4$.
 On the field theory side we find a one-loop Coleman-Weinberg instability in the effective potential for the fundamental scalars.
 We identify a mesonic operator as the dual of the open string tachyon.  By AdS/CFT, we predict the tachyon mass for
 small 't~Hooft coupling (large bulk curvature) and confirm  that it violates the AdS stability bound.

 }

\newcommand{\Tr}{\mbox{Tr}}

\newcommand{\OO}{{\cal O}}
\newcommand{\II}{{\cal I}}

\newcommand{\NN}{{\cal N}}
\newcommand{\MM}{{\cal M}}

\newcommand{\JJ}{{\cal J}}

\newcommand{\KK}{{\cal K}}
\newcommand{\LL}{{\cal L}}

\newcommand{\VV}{{\cal V}}
\newcommand{\XX}{{\cal X}}
\newcommand{\gaugeind}[1]{\mathfrak{#1}}

\def\bea{\begin{eqnarray}}
\def\eea{\end{eqnarray}}
\def\be{\begin{equation}}
\def\ee{\end{equation}}
\def\ea{\end{align}}
\def\bse{\begin{subequations}}
\def\ese{\end{subequations}}
\def\1F1{{}_1\!F_1}
\def\2F0{{}_2\!F_0}


\begin{document}

\section{Introduction}

Open string tachyon condensation has been studied from many viewpoints, see \cite{Sen:2004nf} for a review.
Here we consider a holographic (AdS/CFT) setup where the bulk theory contains an open string tachyon, and ask
what is the counterpart of  tachyon condensation in the boundary field theory.  We will identify
a sector of the boundary theory as a ``holographic open string field theory'' capturing the tachyon dynamics.
Since the bulk  is weakly coupled when the boundary  is strongly coupled, and viceversa,
we are bound to learn something new from their comparison. 

We introduce the open string tachyon by adding to the  $AdS_5 \times S^5$ background two probe $D7$ branes intersecting at general angles. 
Probe branes are the familiar way to include  a small number of fundamental flavors in the AdS/CFT correspondence \cite{Karch:2002sh}.
 If the closed string  background is supersymmetric,
it is possible, and often desirable, to consider configurations of probe branes that preserve some supersymmetry, as {\it e.g.} in \cite{Karch:2002sh, Sakai:2003wu, Kuperstein:2004hy,  
Arean:2004mm, Ouyang:2003df, Canoura:2006es, Apreda:2006bu, Sieg:2007by, Penati:2007vj, Wang:2003yc, Nunez:2003cf, Karch:2000gx, DeWolfe:2001pq, Erdmenger:2002ex, 
Skenderis:2002vf, Constable:2002xt}.
Instead, we are after supersymmetry breaking and the ensuing tachyonic instability. Another way to motivate
our  work is then as  a natural susy-breaking generalization of  the  standard supersymmetric  setup of \cite{Karch:2002sh}.
This generalization is technically challenging, and  the technical aspects have some interest of their own.
 Intersecting brane systems have many other applications in string theory, from string phenomenology to string cosmology, and
 the technical lessons learnt in our problem may be useful in those  contexts as well.

The system that we study is  as an open string analogue of the AdS/CFT
pairs involving closed string tachyons considered in \cite{Adams:2001jb, Dymarsky:2005nc, Dymarsky:2005uh, Pomoni:2008de}. Let us briefly review that analysis.
In all non-supersymmetric orbifolds of ${\cal N}= 4$ SYM,  there is an instability  for large $N$
and small  't Hooft coupling $\lambda$. The instability is
triggered by the renormalization of  double-trace couplings, of the form $f \int d^4 x \;{\cal O}^2$,
where ${\cal O} \sim {\rm Tr} X^2$ is a scalar bilinear.  At leading order for large $N$,
the 't Hooft coupling $\lambda$ is exactly marginal,  but the double-trace coupling  $f$ runs.
The one-loop beta function  $\beta_f (f, \lambda) \equiv \mu \frac{\partial f}{\partial \mu}$ 
does not admit  zeros for real values of $f$  \cite{Dymarsky:2005nc, Dymarsky:2005uh}, so conformal invariance is inevitably broken for arbitrarily small (but non-zero) $\lambda$.
On the field theory side, the instability can be seen in two equivalent ways.  The most
direct is as the Coleman-Weinberg instability of the double-trace part
of the scalar potential, implying that the scalars $X$ must acquire a non-zero vev. Alternatively \cite{Pomoni:2008de},
 we can insist
in formally preserving conformal invariance by tuning $f$ to the zero of its beta function, which is a complex number;
 it then turns out the anomalous dimension $\Delta$ of ${\cal O}$ takes a complex value of the form $\Delta=2 + i b \lambda + O(\lambda^2)$.
By AdS/CFT,  the bulk scalar field dual  to ${\cal O}$ has $m^2  = \Delta (\Delta-4) = -4-b^2 \lambda^2 + O(\lambda^3)$
(in AdS units), and is thus a ``true''  tachyon, since its squared mass is below   the Breitenlohner-Freedman \cite{Breitenlohner:1982bm} 
stability bound $m^2_{BF} = -4$.
The field theory  analysis holds for small 't Hooft  coupling $\lambda$, when 
the bulk string background is strongly curved and the direct evaluation of its spectrum difficult. By contrast
for  $\lambda \to \infty $ calculations are easy in the bulk. The bulk analysis reveals that in some cases (orbifolds with fixed points on $S^5$)
the tachyonic instability persists at large $\lambda$, but it disappears  
 in others (freely acting orbifolds).  The upshot is that AdS/CFT
makes interesting predictions both at weak and a strong coupling. 

The $AdS_5 \times S^5$ background  with two probe intersecting $D7$s can be viewed as
an open string version of this story. For general angles the bulk theory is unstable via condensation of
an open string tachyon, or at least this is the picture for large $\lambda$  where 
we can calculate the string spectrum. In this paper we focus on the  the field theory analysis at small $\lambda$,
with the goal of detecting the expected instability.  

The first challenge is to write down the Lagrangian
of the dual field theory. As is well-known, adding $N_f$ parallel $D7$ branes to $AdS_5 \times S^5$ corresponds
to adding  to  ${\cal N} =4$  SYM action $N_f$ extra ${\cal N}=2$ hyper multiplet in the fundamental representation of the $SU(N)$ gauge group.
The resulting action preserves an ${\cal N}=2$ subalgebra of the original ${\cal N}=4$ supersymmetry algebra -- which
particular ${\cal N}=2$   being a matter of convention so long as it is the same for all the hyper multiplets.
Introducing relative angles between the $D7$ branes corresponds to choosing {\it different} embeddings for  the ${\cal N}=2$ subalgebras
of each different hyper multiplet. In general supersymmetry will be completely broken, while for special angles  ${\cal N}=1$
 susy is preserved. When ${\cal N} =1$ is preserved we can use ${\cal N} =1$ superspace
 to write the Lagrangian. When supersymmetry is completely broken the determination
 of the Lagrangian turns out to be a difficult technical
problem that we are unable to solve completely. We cannot fix the quartic  terms $\sim Q^4$
where $Q$ are the hyper multiplet scalars. The difficulty is related to the lack of an off-shell superspace formulation of ${\cal N}=4$ SYM.
Nevertheless, by making what we believe is a mild technical assumption, we can fix the {\it sign} of the classical quartic potential.
This is sufficient to argue that the theory is indeed unstable from the renormalization
of  ``double-trace'' terms $f \int d^4 x \; {\cal O}^2$, where now ${\cal O} \sim \bar Q^\gaugeind{a} Q_\gaugeind{a}$ with $\gaugeind{a}=1, \dots N$ a color index. 
We are now using ``double-trace'' in quotes since of course
 the fields $Q$ are not matrices but vectors, but the logic is much the same.  The renormalization of $f$ has the same
 twofold interpretation as above. We identify the mesonic operator ${\cal O}$ as the dual of the open string tachyon
 between the two $D7$ branes. The Coleman-Weinberg potential for $Q$ plays the role a holographic effective action for the tachyon.

\section{AdS/CFT with  Flavor Branes Intersecting at General Angles}

We begin with a  review  the Karch-Katz setup \cite{Karch:2002sh}, 
where  parallel  probe $D7$ branes are used to engineer  an $\NN = 2$ supersymmetric field theory with flavor. 
We then break supersymmetry by introducing a relative angle between the $D7$ branes. We derive the dual Lagrangian,
up to an ambiguity in the quartic potential for the fundamental scalars.  We end the section with a review of the
 basic bulk-to-boundary dictionary.

 \begin{figure}[h]
\begin{center}
  \includegraphics[height=4cm,angle=0]{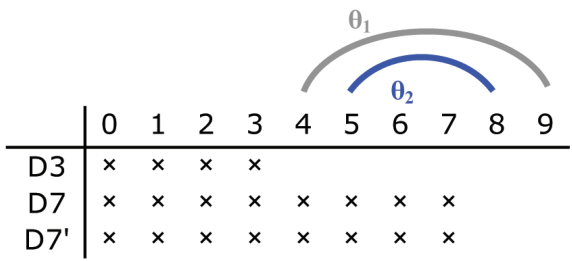}\\
  \caption{\it The brane configuration.}
  \end{center}
\end{figure}

\subsection{Parallel flavor branes}

We start with the familiar $D3/D7$ supersymmetric brane configuration 
 with  $N$ ``color''  $D3$s and $N_f$ ``flavor'' $D7$s, arranged as shown in Figure 1. For
  now $\theta_1 = \theta_2 = 0$, that is,  all $D7$ branes are parallel to one another.
   Taking the decoupling limit on the $D3$s wordvolume, the $D3$ branes are replaced by their near-horizon geometry.
 If $N_f \ll N$, we can treat the $D7$ branes as probes in the  $AdS_5 \times S^5$ background, neglecting their backreaction \cite{Karch:2002sh}.
This background preserves ${\cal N} = 2$ supersymmetry in four dimensions.

The dual field theory is   $\NN=4$  $SU(N)$ SYM coupled to $N_f$ $\NN = 2$ hyper multiplets in the fundamental representation of the $SU(N)$ color group,
arising from the $D3$-$D7$ open strings. We are interested in the case of massless hyper multiplets, corresponding to the brane setup where the $D7$s
coincide with the $D3$s (at the origin of the 89 plane). After decoupling,
the probe $D7$s fill the whole $AdS_5$ and wrap an $S^3 \subset S^5$.

Let us briefly recall the  field content of the boundary theory. A more detailed treatment and the full Lagrangian can be found in Appendix A.
The  $\NN=4$  vector multiplet consists of the gauge field $A_\mu$, four Weyl spinors $\lambda^A_\alpha$, $A = 1, \dots, 4$
 and six real scalars $X_{m}$, $m=4, \dots, 9$ corresponding to the six transverse directions to the $D3$ branes. It
 is convenient to represent the scalars as a self-dual antisymmetric tensor $X^{AB}$  of the $R$-symmetry group $SU(4)_R \cong Spin(6)$,
 \be \label{selfduality}
(X^{AB})^\dagger = \bar{X}_{AB} \equiv \frac{1}{2}\epsilon_{ABCD}\,X^{CD} \,.
\ee
 The explicit change of variables is 
 \be
\label{Xmatrix}
 X^{AB}  = \frac{1}{\sqrt{2}}  \left(\begin{array}{cc|cc}
 0 & X_8 + iX_9 & X_6 + iX_7 & X_4 + iX_5
 \\
 -X_8 - iX_9 & 0 & X_4 - iX_5 & -X_6 + iX_7
 \\
\hline
 -X_6 - iX_7 & -X_4 + iX_5 & 0 & X_8 - iX_9
\\
 -X_4 - iX_5 & X_6 - iX_7 &-X_8 + iX_9 & 0
\end{array}\right)  \,.
\ee
Each ${\cal N} =2$ flavor hyper multiplet consists of two Weyl spinors and two complex scalars,
\be
\begin{array}{ccc}  & \psi^i_\alpha &
\\   q^i  &  &  \left( \tilde{q}_i \right)^\dagger
\\  &  \left( \tilde{\psi}_{i\,\alpha}  \right)^\dagger &
\end{array} 
\ee 
Here $i = 1, \dots, N_f$ is the flavor index. The scalars form an $SU(2)_R$ doublet,
\be
 Q^{\II} \equiv \left(\begin{array}{c} q
\\
\tilde{q}^\dagger
\end{array}\right)   \, , \quad {\II} = 1,2 \,.
\ee    The flavor hyper multiplets
are minimally coupled to the ${\cal N}=2$ vector multiplet that sits
inside the ${\cal N} = 4$ vector multiplet. This coupling breaks the $R$-symmetry
$SU(4)_R$ to $SU(2)_L \times SU(2)_R \times U(1)_R$, where $SU(2)_R \times U(1)_R$
is  the $R$-symmetry of the resulting ${\cal N} = 2$ theory. 
There is a certain arbitrariness
in the choice of embedding $SU(2)_L \times SU(2)_R \times U(1)_R \subset SU(4)_R \cong Spin(6)$.
This corresponds to the choice of 
orientation of the whole stack of  $D7$ branes in the 456789 directions
(we need to pick an ${\mathbb R}^4 \subset {\mathbb R}^6$).
For example if we choose  the configuration of  Figure 1, we identify
$SU(2)_L \times SU(2)_R \cong SO(4)$ with rotations in the $4567$ directions and $U(1)_R \cong SO(2)$ with a rotation on the $89$ plane.
A short calculation using
  our parametrization of the scalars (\ref{Xmatrix}) shows that this
  corresponds to
the following natural embedding of $SU(2)_L \times SU(2)_R \times U(1)_R \subset SU(4)_R$:
\be \label{choice}
\begin{array}{c} 1
\\  2
\\  3
\\  4;p[
\end{array} \,
\left(\begin{array}{cc|cc} SU_R(2) \times U(1)_R & & &
\\  & & & \\
\hline   & & &
\\  & & & SU_L(2) \times  U(1)_R^*
\end{array}\right) \,.
\ee
Of course, any other choice would be equivalent, so long as it is performed simultaneously for all $D7$ branes.
With the choice (\ref{choice}), the $\NN = 4$ vector multiplet splits into
the $\NN = 2$ {vector multiplet} 
\be
\begin{array}{ccc}  & A_\mu &
\\  \lambda^1_\alpha&  & \lambda^2_\alpha
\\  & \frac{X_8 + i X_9}{\sqrt{2}} &
\end{array} \, ,
\ee
and the
 $\NN = 2$ {hyper multiplet}
 \be
\begin{array}{ccc}  & \lambda^3_\alpha &
\\     \frac{X_4 + i X_5}{\sqrt{2}}  &  &  \frac{X_6 + i X_7}{\sqrt{2}}  
\\  & \lambda^4_\alpha &
\end{array}  \,.
\ee 
The two Weyl spinors in the vector multiplet form an $SU(2)_R$ doublet
\be
 \Lambda_{\II} \equiv \left(\begin{array}{c} \lambda_1
\\
\lambda_2
\end{array}\right)  \, , \quad \II = 1,2 \,,
\ee
while the  two spinors in the hyper multiplet form an $SU(2)_L$ doublet,
\be
\hat{\Lambda}_{\hat \II} \equiv \left(\begin{array}{c} \lambda_3
\\
\lambda_4
\end{array}\right) \, , \quad \hat{\II} = 1,2 \,.
\ee  
We use $\II\,, \JJ\, \dots =1,2$ for  $SU(2)_R$ indices and $\hat \II \,, \hat \JJ\, \dots= 1,2$ for $SU(2)_L$ indices.
To make the $SU(2)_L \times SU(2)_R$ quantum numbers of the scalars more transparent we also introduce
the $ 2 \times 2 $ complex matrix $\XX_{\II  \hat \II}$, defined as the off-diagonal block of $X^{A B}$,
 \be \label{calX}
 \XX^{\hat \II \II }  =   \left(\begin{array}{cc}
 X_6 + iX_7 & X_4 + iX_5
 \\
 X_4 - iX_5 & -X_6 + iX_7
\end{array}\right)  \,.
\ee
Note that $ \XX^{\hat \II \II } $ obeys the reality condition
\be
\left( \XX^{\hat \II \II }  \right)^{*}= - \XX_{\hat \II \II } = -  \epsilon_{\hat \II \hat\JJ}  \epsilon_{\II \JJ} \XX^{\hat \JJ  \JJ}   \, .
\ee
We summarize in the following table the transformation properties of the fields:
\begin{table}[h]
\begin{center}
\label{table}
\begin{tabular}{c|c|c|c|c|c}
& $SU(N)$ & $SU(N_f)$   & $SU(2)_L$  & $SU(2)_R$ & $U(1)_R$\\
\hline $A_{\mu}$ & Adj  & $\bf{1}$ & $\bf{1}$ &  $\bf{1}$  & 0
$\vphantom{\raisebox{3pt}{asymm}}$\\
$X^{12}$ & Adj    & $\bf{1}$ & $\bf{1}$ & $\bf{1}$  & +2
$\vphantom{\raisebox{3pt}{asymm}}$\\
$\XX^{\II \hat \II} $   & Adj    & $\bf{1}$  & {\bf 2} & $\bf{2}$ & 0
$\vphantom{\raisebox{3pt}{asymm}}$\\
$\Lambda_{\II}$  &Adj    & $\bf{1}$   & {\bf 1} & $\bf{2}$ & +1
$\vphantom{\raisebox{3pt}{asymm}}$\\
$\hat{\Lambda}_{\hat\II}$  &Adj    & $\bf{1}$  & {\bf 2} & $\bf{1}$ & --1
$\vphantom{\raisebox{3pt}{asymm}}$\\
$Q^\II$  & $\Box$  & $\Box$ & {\bf 1} & $\bf{2}$ & 0
$\vphantom{\raisebox{3pt}{asymm}}$\\
$\psi$ & $\Box$  & $\Box$  & {\bf 1} & $\bf{1}$ & --1
$\vphantom{\raisebox{3pt}{asymm}}$\\
$\tilde{\psi}$ & $\overline{\Box}$  & $\overline{\Box}$  & {\bf 1} & $\bf{1}$ & +1
$\vphantom{\raisebox{3pt}{}}$\\
\end{tabular}
\end{center}
  \caption{\it Quantum numbers of the  fields.  }
\end{table}

\subsection{Rotating the flavor branes \label{rotate}}

We now describe a non-supersymmetric open string deformation of this background.
 For simplicity we consider the case $N_f=2$. While keeping the two $D7$ branes coincident with the $D3$s  in the 0123 directions,
we rotate them with respect to each other in the transverse six directions, see Figure 1.
 There are two independent angles, so without loss of generality we may
perform a rotation of angle $\theta_1 = \theta_{49}$ in the 49 plane and  a rotation of angle $\theta_2 = \theta_{85}$ in the 58 plane. For generic angles
supersymmetry is completely broken; for  $\theta_1= \theta_2 $  it is broken to  $\NN = 1$.
As we rotate the branes, some  $D7$-$D7'$ open string modes become tachyonic. 
The main goal of this paper is to study this tachyonic instability from the viewpoint of the dual field theory.

On the field theory side, rotating the second brane $D7'$ amounts to choosing a different embedding of $SU(2)_R \subset SU(4)$ 
for the second hyper multiplet, while keeping the standard embedding (\ref{choice}) for the first.
In the Lagrangian, we must perform an $SU(4)$ rotation of the $\NN=4$ fields 
 that couple to the second hyper multiplet, leaving the ones that couple to the first  unchanged. The rotation is
 of the form
 \be \label{rotation}
 X'_m = {\cal R}^{ ({\bf 6} ) \;n}_{\;m} (\theta_{1}, \theta_{2})  \, X_n \, , \qquad  \lambda'_A = {\cal R}^{ ({\bf 4}) \; B}_{\;A} (\theta_{1}, \theta_{2}) \,   \lambda_B\, .
 \ee
The explicit form of the rotation matrices  $ {\cal R}^{ ({\bf 6} )}$  and $ {\cal R}^{ ({\bf 4} )}$ is given in Appendix B.

 Naively, the $Q^4$ terms are not affected by the rotation, but this is incorrect. This is seen clearly in $\NN=1$ superspace.
 The $\NN=4$ multiplet is built out of three chiral multiplets $\Phi^a$, $a=1,2,3$ and one vector multiplet $V$.
 The $Q^4$ terms arise from integrating out the auxiliary fields $F^a$ ($a=1,2,3$) of the chiral multiplets and $D$ of the vector multiplet,
 which transform  under the $SU(4)_R$ rotation.  For example, a rotation that preserves
 $\NN =1$ supersymmetry ($\theta_1 = \theta_2$) corresponds to a matrix ${\cal R}^{({\bf 4})} \subset SU(3)$,
 which acts on  $F^a$ leaving $D$ invariant. The correct Lagrangian is obtained by performing
 the rotation on the $X_m$, $\lambda_A$ {\it and}  $F^a$   fields that couple to the primed hyper multiplet,
 and only then can the auxiliary fields be integrated out.  The $Q^4$ terms get modified accordingly.

Under a more general $SU(4)_R$ rotation, the $F^a$ and $D$ auxiliary fields  are expected to mix in a non-trivial fashion. 
 There exists a formalism developed in \cite{Marcus:1983wb, Gates:1983nr}
that provides  the generic R-symmetry 
transformations action in $\NN =1$ superspace.
Unfortunately, for $\NN=4$ supersymmetry we cannot rely on this formalism because the transformations do not close off-shell.
This  technically involved point  is explained in detail in Appendix C. There we also provide an $\NN=2$ supersymmetry toy example where the formalism works perfectly since the $\NN=2$ R-symmetry algebra closes off-shell.

To proceed, we parametrize our ignorance of the $Q^4$ terms. The exact form of the full Lagrangian, including the parametrized $Q^4$ potential, is spelled out in Appendix B.
Schematically, we write the $Q^4$ potential as
\be
 V_{Q^4} =  Q^4_1  +  Q^4_2 +   \left(Q_1Q_2 \right)^2_F \,f\left( \theta_1 , \theta_2 \right)  +   \left(Q_1Q_2 \right)^2_D \,d\left( \theta_1 , \theta_2 \right) \, ,
\ee
 for some unknown functions $f(\theta_1, \theta_2)$ and $d(\theta_1, \theta_2)$. Here $Q_1$ and $Q_2$ are shorthands for the scalars in the first and second hyper multiplets
and the subscripts $F$ and $D$ refer to different ways to contract the indices, see (\ref{the potential}) for the exact expressions. The letters $F$ and $D$ are chosen as reminders
of the (naive) origin of the two structures from integrating out the ``rotated'' $F$ and $D$ ${\cal N}=1$ auxiliary fields,
but  this form of the potential  follows from rather general symmetry considerations, as we explain in Appendix B.
When $\theta_1 = \theta_2$, ${\cal N}=1$ supersymmetry is preserved and ${\cal N}= 1$ superspace allows to fix the two functions,
\be
f(\theta, \theta) = \cos \theta \, ,\qquad d(\theta, \theta) =1 \,.
\ee
For general angles, we can constrain $f$ and $d$ somewhat, using bosonic symmetries (see Appendix B),
but unfortunately we are unable to fix them uniquely.
The most important assumption we will make in the following is {\it positivity} of the classical potential,  $V_{Q^4} \geq 0$,
implying $f(\theta_1, \theta_2) \leq 1$ and $d(\theta_1, \theta_2) \leq 1$ for all $\theta_1$, $\theta_2$.
Positivity would follow from the mere existence of any reasonable off-shell superspace formulation,
as the scalar potential  would always be proportional to the square of the auxiliary fields,
 even when supersymmetry
is broken by the relative R-charge rotation between the two hyper multiplets.\footnote{To illustrate how this would work we consider 
in section C.2  ${\cal N}=2$ SYM theory coupled to two fundamental ${\cal N}=1$
chiral multiplets, with different choices of the two ${\cal N}=1$ subalgebras.}
Note also that the classical potential $V_{Q^4}$ is a homogeneous function of the $Q$s, so it is everywhere positive
if and only if it is bounded from below, which is another plausible  requirement.

\subsection{Bulk-boundary dictionary}

The basic bulk-to-boundary dictionary for the parallel brane case has been worked
out in \cite{Aharony:1998xz, Kruczenski:2003be}. A brief review is in order.

 In the closed string sector, Type IIB closed string fields map to
 single-trace operators of ${\cal N} = 4$ SYM, as usual. In the open string sector, open string fields on the $D7$ worldvolume 
map to gauge-singlet mesonic operators, of the schematic form $\bar Q X^n Q$, where $Q$ stands for a generic fundamental field and $X$ for a generic adjoint field.

The massless bosonic fields  on the $D7$ worldvolume are a scalar $\Phi$
and a gauge field $(A_{\hat{\mu}} , A_{\hat{\alpha}})$,  where $\hat{\mu}$ are $AdS_5$  indices and $\hat{\alpha}$  are $S^3$ indices.
Kaluza Klein reduction on the $S^3$ generates the following tower of states, labeled in terms of  $\left(j_1\,,\,j_2 \right)_s$ representations of $SU(2)_L \times SU(2)_R \times U(1)_R$:
\be
\Phi \rightarrow  \Phi^\ell = \left(\frac{\ell}{2}\,,\,\frac{\ell}{2} \right)_2  \, ,\quad A_{\hat{\mu}}   \rightarrow  A_{\hat{\mu}}^\ell = \left(\frac{\ell}{2}\,,\,\frac{\ell}{2} \right)_0    \, ,\quad A_{\hat{\alpha}}   \rightarrow  A_{\pm}^\ell = \left(\frac{\ell \pm 1}{2},\,\frac{\ell \mp 1}{2} \right)_0 \,.
\ee
(The longitudinal component of $A_{\hat{\alpha}} $ is not included because it can be gauged away). These states (and their fermionic partners, which we omit) can be organized
into short multiplets of the $\NN = 2$ superconformal algebra, 
\be
\left(A_-^{\ell+1},A_{\hat \mu}^\ell,\Phi^\ell,A_+^{\ell-1} \right) \,, \quad \ell = 0,1,2, \dots
\ee
of conformal dimensions
\be
(\ell +2, \ell+3, \ell+3, \ell +4) \,.
\ee
For $\ell = 0$ the $A_+$ state is absent. Note that all states in a given multiplet have the same $SU(2)_L$ spin, indeed the $\NN = 2$ supercharges are neutral under $SU(2)_L$.

The lowest member of each multiplet, namely $A_-^{\ell + 1}$, is dual to the chiral primary operator
\be \label{chiralmeson}
\bar Q_{\{ \II }  \XX_{ \II_1 \, \hat \II_1 } \dots \XX_{ \II_\ell  \, \hat \II_\ell }  Q_{  \JJ  \}  } \, ,
 \ee
 where
 $Q^\II$
 is the $SU(2)_{R}$ doublet of complex fundamental scalars. In (\ref{chiralmeson}) the $SU(2)_L$ and $SU(2)_R$ indices are separately symmetrized. 
In particular for $\ell =0$, we have the triplet of mesonic operators
 \be
\label{triplet}
\OO_{\bf{3}} \equiv   \bar{Q}_{ \{ \II  } Q_{\JJ  \} } =   \bar{Q}_\II  Q^\JJ  -  \frac{1}{2}\bar{Q}_\KK  Q^\KK \delta^\JJ_\II \,.
\ee
The singlet operator
\be
\OO_{\bf{1}} \equiv \bar{Q}_\II  Q^\II \, ,\quad 
\ee
is not a chiral primary and maps to a massive open string state.

\begin{figure}[t]
\begin{center}
  \includegraphics[height=4cm,angle=0]{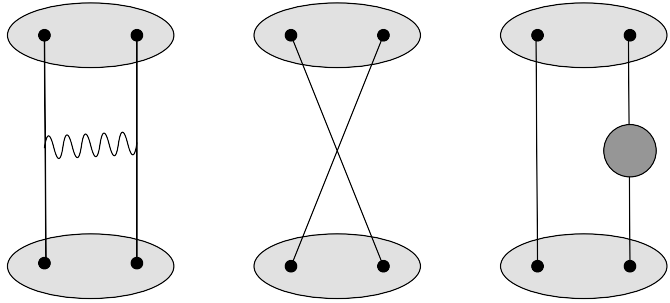}\\
  \caption{\it  Diagrams contributing to the one-loop renormalization of the mesonic operators.}
  \label{renormoper}
  \end{center}
\end{figure} 
In Appendix D we compute the one-loop dilatation operator acting on the basis of states $\OO  \equiv \bar{Q}_\II  Q^\JJ$, evaluating
the diagrams schematically drawn in Figure 2.  We find
 \be
   \Gamma^{(1)} = \frac{\lambda}{4\pi^2} \,\mathbb{K}  \, , \quad \mathbb{K}\equiv \delta^\II_\JJ \delta^\KK_\LL    \,.
   \ee   
 The eigenstates are the triplet and the singlet, with eigenvalues
 \be
 \gamma_{\bf{3}} = 0
\, ,\quad
 \gamma_{\bf{1}} = \frac{\lambda}{2\pi^2} \,.
\ee
As expected, the chiral triplet operator has protected dimension. At one-loop,
this result does not change as we turn on non-zero angles $\theta_1$ and $\theta_2$.

So far we have considered the case of a single $D7$ brane, or a single flavor. For multiple $D7$ branes
(multiple flavors) the Chan-Paton labels  of the open strings are interpreted as the bifundamental flavor indices
of the mesonic operators, ${\cal O}^{ij}$, $i,j=1,\dots N_f$.  In our setup, with $N_f=2$,  the lowest mode of the open string with off-diagonal Chan-Paton labels,
which is the massless gauge field for parallel branes, becomes tachyonic as we turn on a relative angle between the $D7$s.  In the dual field theory
we expect to find an instability  associated with the operator ${\cal O}_{\bf 3}^{12}$,  the lowest dimensional operator dual to the off-diagonal open string mode.

\section{``Double-trace'' Renormalization and the Open String Tachyon}

Our setup is an open string version of the phenomena studied in  \cite{Adams:2001jb, Dymarsky:2005nc, Dymarsky:2005uh, Pomoni:2008de}.  
Motivated by the work of  \cite{Dymarsky:2005nc,Dymarsky:2005uh}, we considered in  \cite{Pomoni:2008de} a generic large $N$, non-supersymmetric field theory with all matter in the adjoint (or bifundamental) representation. We further assumed the theory to be ``conformal in its single-trace sector", by which we mean that all single-trace couplings have vanishing beta function for large $N$. In such a theory, quantum effects induce double-trace couplings of the schematic form
\be
\delta S = f \int d^4x \, \OO \bar{\OO} \, , \quad \OO \sim \Tr\, \phi^2\,,
\ee
where $\phi$ is a scalar field.  The beta function for $f$ may or may not admit a real fixed point. If $\beta_f$ has no real zeros, conformal invariance is broken.
A closely related phenomenon is the generation of a Coleman-Weinberg potential $\mathcal{V}\left( \langle \phi \rangle \right)$ \cite{Adams:2001jb}. It is not difficult to show \cite{Pomoni:2008de} that the symmetric vacuum $\langle \phi \rangle=0$ is stable if and only if $\beta_f$ has a fixed point; conversely, if $\beta_f$ has no real zeros, dynamical symmetry breaking occurs.

Field theories of the kind just described arise in several examples of the AdS/CFT correspondence. The best known cases are non-supersymmetric orbifolds of $\NN=4$ SYM, dual to Type IIB string theory on $AdS_5\times S^5/\Gamma$ with $\Gamma $ a subgroup of $SU(4)_R$ (but not a subgroup of $SU(3)$). 
The field theory instability associated with double-trace renormalization is the boundary counterpart of the instability associated with a closed string tachyon in the AdS bulk. By a tachyon we mean a bulk scalar that {\it violates} the Breinlohner-Freedman bound. As usual, the weakly coupled boundary theory (small 't Hooft coupling $\lambda$) gives information about the high curvature regime of the bulk theory and vice versa. In some examples (non-freely acting orbifolds of $\NN=4$ SYM), the instability is visible both in the weakly curved  bulk theory and in the one-loop analysis of the boundary theory; presumably the theory is unstable for all couplings. In other examples (freely acting orbifolds of $\NN=4$) an instability shows up in the one-loop analysis of the boundary theory, but the spectrum of the weakly curved bulk theory has no tachyon; the tachyon must become massive for $\lambda$ greater than some critical value.

We showed in \cite{Pomoni:2008de}  that at large $N$ the conformal dimension $\Delta_{\OO}$  and the beta function $\beta_f$ take the general forms
\be \label{DeltaOform}
 \Delta_{{\cal O} } = 2 + \gamma(\lambda) + \frac{ v(\lambda) }{1+ \gamma(\lambda)}f \,  ,
 \ee
\be  \label{betafform}
\beta_f =\frac{ v(\lambda) }{1+ \gamma(\lambda)}f^2 + 2 \gamma(\lambda) f + a(\lambda) \, .
\ee
Here $ v(\lambda)$ is defined as the normalization coefficient of $ \OO$,
\be
\langle \OO(x) \bar\OO(0) \rangle = \frac{ v(\lambda) }{2\pi^2x^{2 \Delta (\lambda)}}   \, \, ;
\ee
$\gamma(\lambda)$ is the contribution to the anomalous dimension of  $\OO$ from single-trace interactions; finally $a(\lambda)$ is the coefficient of the induced 
double-trace terms, coming from the single trace interactions, in the quantum effective potential.
The expressions (\ref{DeltaOform}, \ref{betafform})  are valid to all orders in  planar perturbation theory: large $N$ factorization implies that $\beta_f$ depends at most quadratically on $f$, and $\Delta_{\OO}$ at most linearly  \cite{Pomoni:2008de} .
The coefficients $v(\lambda)$, $\gamma(\lambda)$ and $a(\lambda)$ have planar perturbative expansions
\be
v(\lambda) = \sum_{L=1}^\infty v^{(L)} \lambda^{L-1} \,,\quad \gamma(\lambda) = \sum_{L=1}^\infty \gamma^{(L)} \lambda^L \,,\quad a(\lambda) = \sum_{L=1}^\infty a^{(L)} \lambda^{L+1} \,,
\ee
where the  $L$ denotes the loop order. Consider the discriminant of the quadratic equation $\beta_f=0$,
\be
D (\lambda) \equiv \gamma(\lambda)^2 -   \frac{a(\lambda)\, v(\lambda) }{1+ \gamma(\lambda)} \, .
\ee
If $D(\lambda) <0$,  there are no physical (real) values of $f$ for which the theory is conformal. But if we insist on formally preserving conformal invariance by
 tuning $f$ to one of its two complex fixed points,  
 then the operator dimension also becomes complex,
 \be
 \Delta_{\OO} = 2 \pm  \, i \, b(\lambda)\,, \qquad b(\lambda) \equiv \sqrt{|D|}\,.
 \ee
Using the usual AdS/CFT dictionary 
\be
 \Delta_{\OO} = \frac{d}{2} \pm \sqrt{\frac{d^2}{4}+m^2R^2} = 2 \pm \sqrt{4+m^2R^2}   \, ,
\ee
we find that the $AdS_5$ scalar dual to ${\cal O}$  has mass \cite{Pomoni:2008de}
\be \label{tachyonmass}
m^2(\lambda) R^2  = m^2_{BF} R^2 +D(\lambda)= - 4 +D(\lambda)  \,.
\ee
For negative discriminant $m^2(\lambda)  < m^2_{BF} $: the scalar field dual to $\OO$  is a true tachyon and the bulk theory is unstable.

We now generalize this story 
to AdS/CFT dual pairs containing an open string sector. In the presence of flavor branes in the AdS bulk, the dual field theory contains extra fundamental matter. An {\it open} string tachyon corresponds to an instability in the {\it mesonic} sector of the boundary theory. In this paper we illustrate this phenomenon in the example of the intersecting $D7$ brane system. The classical Lagrangian of the boundary theory takes the schematic form 
\be
\LL = \LL_{adjoint}+ \LL_{fund} =  -\Tr \left[  F^2 + \left(D X \right)^2  +\dots \right]  -\left(D Q \right)^2 - \frac{\lambda}{N}  {\cal O}^{ij} {\bar{\cal O}}^{ij}+ \dots
\ee
where $ {\cal O}^{ij}  = Q^{i \,\gaugeind{a}} \bar Q_{j \, \gaugeind{a}}$ are the gauge-invariant mesonic operators made from the fundamental scalars and for simplicity we have ignored
 the $SU(2)_R$ structure, which will be restored shortly.\footnote{To avoid cluttering in some expressions below we always write the flavor indices as upper  indices in ${\cal O}^{ij}$.}
The whole $ \LL_{fund} $ is $1/N$ suppressed with respect to  $\LL_{adjoint}$, in harmony with the fact that the classical D-brane
effective action arises from worldsheets with disk topology, and is thus suppressed by a power of $g_s\sim 1/N$ with respect to the classical closed string effective action,
arising from worldsheets with sphere topology. Nevertheless, as always
in the tachyon condensation problem, it makes perfect sense to focus on classical open string field theory. The classical open string dynamics 
 is dual to the quantum planar dynamics of the mesonic sector of the field theory.
 The 't Hooft coupling $\lambda$ does not run at leading order in $N$, indeed  the hyper multiplet contribute to $\beta_\lambda$ at order $O(1/N)$.
For generic angles, the term in the $Q^4$ potential that mix the two flavors run, so perturbative renormalizability forces the introduction of a new coupling constant
$f$, 
\be\label{extradt}
\delta \LL_{fund} = -\frac{f}{N}  {\cal O}^{12} {\bar {\cal O}}^{12}\,.
\ee
Note on the other hand that no extra terms diagonal in flavor (namely ${\cal O}^{11} {\bar {\cal O}}^{11}$ and ${\cal O}^{22} {\bar {\cal O}}^{22}$
are induced at one-loop. For the first flavor  this is immediate to see: the diagrams contributing to the term  ${\cal O}^{11} {\bar {\cal O}}^{11}$  of the effective
potential are independent of $\theta_1$, $\theta_2$ (they do not involved any coupling mixing the two flavors) 
and thus their sum must vanish, as it does in the ${\cal N}=2$ supersymmetric theory with $\theta_1 = \theta_2 =0$.
For the second flavor this follows by symmetry, since the two flavors are of course interchangeable.\footnote{In more detail, the diagrams
contributing to ${\cal O}^{22} {\bar {\cal O}}^{22}$  do not involve the first flavor, which could then be set to zero as the calculation of this terms
of the effective potential
is concerned. The Lagrangian with the first flavor set to zero is ${\cal N}=2$ supersymmetric, only with an unconventional choice of $SU(2)_R$ embedding into $SU(4)_R$ --
it can be turned into the standard Lagrangian by an R-symmetry rotation of the ${\cal N}=4$ fields.}

This extra ``double-trace'' term (\ref{extradt}) arise at
  the same order in $N$ order as the classical $\LL_{fund}$, indeed  inspection of the Feynman diagrams shows that the one-loop bare coupling $f_0$ behaves as
  \be
f_0 \sim \lambda^2 \log \Lambda \,.
\ee
The analysis \cite{Pomoni:2008de}  can be applied  in its entirety to this ``open string'' case. The ``double-trace'' beta function $\beta_f$ takes again the  form (\ref{betafform}),
and  its discriminant $D(\lambda)$ computes now (through (\ref{tachyonmass})  the mass of the {\it open} string tachyon dual to mesonic operator ${\cal O}^{12}$.
Let us turn to explicit calculations.

\subsection{The one-loop ``double-trace'' beta function}

To proceed, we need to be more precise about the structure of the``double-trace'' terms induced at one-loop, restoring their $SU(2)_R$ structure.
For general angles $\theta_1$ and $\theta_2$, there are three independent structures,
\be
\label{deformation}
\delta \LL_{fund} =  -\frac{1}{N} \left[ f_{  {\bf 3}^\pm  }  \left(  {\cal O}_{  {\bf 3}^+}^{12}  {\cal O}^{21}_{{\bf 3}^-} + {\cal O}_{{\bf 3}^-}^{12} { {\cal O}}^{21}_{{\bf 3}^+}  \right) + 
 f_{  {\bf 3}^0 } {\cal O}_{{\bf 3}^0}^{12} { {\cal O}}^{21}_{{\bf 3}^0}  +   f_{  {\bf 1}}{\cal O}_{{\bf 1} }^{12} { {\cal O}}^{21}_{{\bf 1}}   \right] \,.
\ee
We have imposed neutrality  under the Cartan of $SU(2)_R$, since this is an exact symmetry for generic angles, corresponding geometrically
to rotations in the 67 plane (more precisely a 67 rotation is a linear combination of the Cartan of $SU(2)_L$ and $SU(2)_R$, but the hyper multiplets
are neutral under $SU(2)_L$). When one of the angles is zero, say $\theta_2=0$, rotations in the 567 directions a symmetry (again a diagonal combination of $SU(2)_L$ and $SU(2)_R$),
implying $f_{{\bf 3}^\pm} = f_{{\bf 3}^0} \equiv f_{\bf 3}$. We  focus on the triplet mesons, which are dual to the open string tachyon.  For a single non-zero angle
there is one beta function  $\beta_{f_{{\bf 3}}}$ to compute,  since the three components of the triplet are related by symmetry. For generic angles there are in principle two distinct
beta functions $\beta_{f_{{\bf 3}^\pm}}$ and $\beta_{f_{{\bf 3}^0}}$; we will illustrate our method computing the first, which is a slightly simpler calculation.
At one-loop,  the ``double-trace'' beta function takes the form 
\be
\beta_f = v^{(1)} f^2 + 2 \gamma^{(1)} \lambda  f + a^{(1)}\lambda^2 \,.
\ee
We have seen that $\gamma^{(1)} = 0$ at one loop for the triplet mesons. The normalization coefficient $v^{(1)}$ is easily evaluated
by free Wick contractions,
\be
\langle \OO^{12}_{\bf{3}^0}\,(x) \OO^{21}_{\bf{3}^0}\,(y) \rangle = \langle \OO^{12}_{\bf{3}^+}\,(x) \OO^{21}_{\bf{3}^-}\,(y) \rangle = \langle \OO^{12}_{\bf{3}^-}\,(x) \OO^{21}_{\bf{3}^+}\,(y) \rangle = \frac{1}{16 \pi^4 |x-y|^4} \, ,  
\ee
implying
\be
 v_{\bf{3}^+}^{(1)}= v_{\bf{3}^-}^{(1)}= v_{\bf{3}^0}^{(1)}=\frac{1}{8\pi^2}\,.
\ee
It remains to evaluate the coefficient $a^{(1)}$. We are going to extract  $a^{(1)}$ from the one-loop
 Coleman-Weinberg potential along the ``Higgs branch'' of the gauge theory, $\langle X_{A B} \rangle = 0$, ${\cal Q}^{\cal I} \neq 0$. We put ``Higgs branch'' in quotes because
 for general angles it is in fact lifted already at the classical level. Let us first recall the analysis 
  for the ${\cal N} =2$ supersymmetric theory corresponding to two parallel flavor branes are parallel ($\theta_1 = \theta_2 =0$).

As always in a  supersymmetric theory,  flat directions are parametrized by holomorphic gauge-invariant composite operators. In our case the relevant operators are the mesons
\be
\label{mesons}
\OO^{ij} =q^i\cdot \tilde{q}_j\, ,\quad i,j=1,2
\ee
The dot stands for color contraction $q \cdot {q}^* \equiv q^\gaugeind{a}\,q^*_\gaugeind{a}  $ and $i,j$  are the flavor indices.
The  holomorphic, gauge invariant mesons that parameterize the Higgs flat directions are subject to F-flatness conditions
\be
q^{\gaugeind{a}\,i} \tilde{q}_{\gaugeind{b}\,i} =0 \Leftrightarrow  \, {\rm tr}\, {\OO} = {\rm det}\, {\OO} =0\, ,
\ee
thus there are $4-2=2$ complex parameters for the moduli space of the supersymmetric theory ($\theta_1 = \theta_2=0$). We may parameterize the flat directions by
\be
\label{unrotated flat}
Q_1 = U  \left(\begin{array}{c} q
\\
0
\end{array}\right)   
\, ,\quad
Q_2 = U  \left(\begin{array}{c} 0
\\
-q
\end{array}\right) 
\, ,\quad
U \in SU(2) \quad \mbox{and} \quad q\in \mathbb{R} \, .
\ee
Color indices are kept implicit. In color space we may take $q^{\gaugeind{a}=1}=q$ and   $q^{\gaugeind{a} \neq 1}=0$.
For generic $\theta_1$, $\theta_2$ supersymmetry is explicitly broken in the classical Lagrangian and the Higgs branch is completely lifted.

 To select $\beta_{f_{{\bf 3}^+}}$ (which is of course equal to $\beta_{f_{{\bf 3}^-}}$),  we calculate the effective
 potential around a classical background such that $\langle {\cal O}_{\bf 1}^{12} \rangle   =\langle {\cal O}_{{\bf 3}^0}^{12} \rangle   = \langle {\cal O}_{{\bf 3}^-}^{12} \rangle= 0$, but $\langle {\cal O}_{{\bf 3}^+}^{12} \rangle \neq 0$,
 namely
\be
\label{back}
Q_1 =   \left(\begin{array}{c} q
\\
0
\end{array}\right)   
\, ,\quad
Q_2 =   \left(\begin{array}{c} 0
\\
-q
\end{array}\right) 
\, ,\quad
 \quad q\in \mathbb{C}
\ee
This choice  corresponds to the 
 flat direction for the ${\cal N}=1$ susy case $\theta_1 =\theta_2$.
 The F-terms of the classical potential vanish for general angles, but for $\theta_1 \neq \theta_2 $ the D-terms do not,   ${\cal V}^D_{Q^4} =g^2 |q|^4 ( 1-  d\left(\theta_1,\theta_2 \right) )$.

 In Appendix E we evaluate the one-loop contribution to the effective potential along this background (at large $N$).
With the help of the Callan-Symanzik equation we find
\be
a^{(1)}_{{\bf 3}^{\pm}}= \frac{1}{16\pi^2} \left[  \Big(1-d(\theta_1 , \theta_2)  \Big) 
 + \frac{1}{2} \Big(1-d(\theta_1 , \theta_2)  \Big)^2
+ 4\,\sin^2{\left(\frac{\theta_1 + \theta_2}{2} \right)}  \sin^2{\left(\frac{\theta_1 - \theta_2}{2} \right)}\, \right] \,.
\ee
From our (mild) assumption that the classical potential be positive we have $d(\theta_1, \theta_2) \leq 1$, has the crucial implication 
\be
a^{(1)}_{{\bf 3}^{\pm}} \geq 0\,.
\ee
In the supersymmetric case ($\theta_1=\theta_2$), $a^{(1)}_{{\bf 3}^{\pm}} = 0$, as it must. For $\theta_2 =0$, the $SU(2)$ symmetry is restored, so 
\be
a^{(1)}_{{\bf 3}^{\pm}}= a^{(1)}_{{\bf 3}^0}=
\frac{1}{16\pi^2} \left[  \Big(1-d(\theta , 0)  \Big) 
 + \frac{1}{2} \Big(1-d(\theta , 0)  \Big)^2
+ 4\,\sin^4{\left(\frac{\theta }{2} \right)}  \, \right] \geq 0\,.
\ee
The one-loop triplet beta function (let us focus on the single-angle case)
\be
\label{beta3}
\beta_{f_{\bf 3}} = v^{(1)}_{\bf{3}} \,f^2_{\bf 3} + a^{(1)}_{\bf 3} \lambda^2
\ee
does not admit real fixed points for $f_{\bf 3}$, so conformal
invariance is inevitably broken in the quantum theory.\footnote{Conformal invariance is already broken
in the adjoint (``closed string'') sector  by the hyper multiplet contribution
to $\beta_\lambda$, but this is subleading effect (of order $O(1/N)$) with respect to the classical Lagrangian.
 In the fundamental (``open string'') sector the breaking of conformal
invariance is at leading order in $N$ (quantum effects arise as the same order as the classical Lagrangian).
Of course the {\it whole} fundamental sector is $O(1/N)$ with respect to the adjoint sector, but we can meaningfully separate the effect
we are interested in. This is the field theory counterpart of focussing on the classical open string dynamics of the D-branes, while
ignoring the backreaction of the branes on the bulk background.}
The running coupling 
\be
\bar f(\mu) =  \frac{ a^{(1)}} {\sqrt{v^{(1)}_{\bf{3}}}} \, \lambda^2 \, \tan \left[ \frac{a^{(1)}\lambda^2} {\sqrt{v^{(1)}_{\bf{3}}}} \ln (\mu/\mu_0)  \right] 
\ee
is a monotonically increasing function interpolating between IR and UV Landau poles, at
energies
\be
\mu_{IR} = \mu_0 \, \exp\left(-\frac{\pi \sqrt{v^{(1)}_{\bf3}   }   }{    \lambda \sqrt{a^{(1)}_{\bf{3} }  }  } \right) \, , \quad \mu_{UV} = \mu_0 \,\exp \left(\frac{\pi \sqrt{v^{(1)}_{\bf{3}}}}{  \lambda\sqrt{a^{(1)}_{\bf{3}}}} \right) \,. \ee
For small coupling $\lambda \to 0$, the Landau poles are pushed respectively to zero and infinity.

\subsection{The tachyon mass}

As reviewed above, the mass of the  field dual to ${\cal O}^{12}_{{\bf 3}^\pm}$ is
directly related to the discriminant of $\beta_{{\bf 3}^\pm}$,
\be \label{tachyonmass2}
m^2_{{\bf 3}^\pm} R^2 = m^2_{BF}R^2 + D_{{\bf 3}^\pm}(\lambda; \theta_1, \theta_2)=
-\,4\, -\frac{ \lambda^2}{16\pi^4} \, \mathcal{D}^{(1)}_{\bf{3}}\left(\theta_1, \theta_2 \right)+ \OO(\lambda^3) \, ,
\ee
where 
\be 
 \mathcal{D}_{\bf{3}}^{(1)} = \Big(1-d(\theta_1 , \theta_2)  \Big) 
 + \frac{1}{2} \Big(1-d(\theta_1 , \theta_2)  \Big)^2
+ 4\,\sin^2{\left(\frac{\theta_1 + \theta_2}{2} \right)}  \sin^2{\left(\frac{\theta_1 - \theta_2}{2} \right)} \, .
\ee
For $\theta_1 \neq \theta_2$ the discriminant is negative, implying that the bulk
field violates the BF stability bound. 
Whenever  supersymmetry is broken,  the bulk field dual to ${\cal O}^{12}_{{\bf 3}^\pm}$ is a true tachyon. 
For $\theta_2 =0$ the $\pm$ and $0$ components of the triplet are related by the $SU(2)$ symmetry and are all tachyonic.
We expect the field dual to ${\cal O}^{12}_{{\bf 3}^0}$ to be tachyonic for general angles.

For small angles, the $O(\lambda^2)$ tachyon mass depends on a single unknown parameter $\alpha$ (which enters the parametrization of the classical $Q^4$ potential, see Appendix B), 
\be \label{weak}
R^2 m_{\bf 3}^2(\lambda) =-4 - \alpha \, \frac{ \lambda^2}{16\pi^4}\left( \theta_1- \theta_2\right)^2    + \OO(\lambda^3)\, , \qquad \theta_1\, ,\theta_2 \ll 1\,.
\ee
This  expression applies to all three components of the triplet.
For the $\pm$ components it is just the expansion of (\ref{tachyonmass2}) for small angles. For the $0$ component it follows by imposing the symmetry constraints
$m_{{\bf 3}^0}^2(\theta, \theta)=0$ and $m_{{\bf 3}^0}^2(\theta, 0)=m_{{\bf 3}^\pm}^2(\theta, 0)$.  By AdS/CFT, we get an interesting prediction
for the mass of the open string tachyon for large AdS curvature (small $\lambda$). 

Conversely, for large $\lambda$ (small AdS curvature) we can compute the mass of the open string tachyon using the dual string picture. The open string spectrum
of branes intersecting at small angles in flat space is well-known. The lowest tachyon mode has mass
 (see {\it e.g.} \cite{Epple:2003xt} for a review),
 \be
 m^2 = -\frac{| \theta_1- \theta_2|}{\pi^2 \alpha'} \,  , , \qquad \theta_1\, ,\theta_2 \ll 1\,.
 \ee
This  becomes a good approximation to the  mass in the  exact AdS sigma model in the
 limit $\alpha'/R^2 \sim \lambda^{-1/2} \to 0$. Thus
\be \label{strong}
\lim_{\lambda \to \infty} R^2 m_{\bf 3}^2(\lambda) = -\frac{| \theta_1- \theta_2|}{\pi^2} \frac{R^2}{\alpha'} =  -\frac{| \theta_1- \theta_2|}{\pi^2} \lambda^{1/2}\, ,
, \qquad \theta_1\, ,\theta_2 \ll 1\,.
\ee
This can be regarded as a prediction for the large $\lambda$ behavior of the discriminant $D_{\bf 3}(\lambda)$, which is a purely field-theoretic quantity.
Note that apart from the $\lambda$ dependence, which could have been anticipated on general grounds, the weak coupling result (\ref{weak}) and the strong coupling
result (\ref{strong}) differ in their angular dependence.

\section{Discussion}

The main technical question that we leave answered is   the precise form of the classical $Q^4$ potential for generic angles.
As we have emphasized, a superspace formulation of ${\cal N}=4$ SYM with manifest $SU(4)_R$ symmetry
would offer a solution. It would be interesting to see whether the new off-shell  formalism  for $\NN=1$ SYM in ten dimensions 
introduced in \cite{Berkovits:1993zz, Baulieu:2007ew} could be applied to our problem.
 In principle, another way to obtain the $Q^4$ potential is by taking the decoupling limit
of the intersecting brane effective action. This would first require the calculation of a four-point function of twist fields, two twist fields corresponding to  D3-D7 open strings
and two twist fields corresponding to D3-D7' open strings. This problem has been solved for branes intersecting at right angles 
 (see  {\it e.g.} \cite{Dixon:1986qv, Cvetic:2003ch, Abel:2003vv, Antoniadis:2000jv}). The generalization to arbitrary angles is an interesting
 and  difficult problem in boundary conformal field theory. Taking the decoupling limit may also be challenging in the presence
 of tachyons -- it is not clear to us whether the result would be unambiguous or it would require some renormalization prescription.

Even without a complete knowledge  of the classical $Q^4$ potential, by making a plausible positivity assumption
we argued that the field theory is unstable at the quantum level. By AdS/CFT, we obtained a non-trivial
 prediction for the tachyon squared mass $m_{\bf 3}^2(\lambda)$ at small $\lambda$. Its behavior at large $\lambda$ is known
 from flat-space string theory. There must exist an
 interpolating function $m_{\bf 3}^2(\lambda)$ valid for all $\lambda$. It would be extremely
 interesting to apply integrability techniques to find the whole function. There is a large
 literature on  open  spin chains arising in the calculation of anomalous
 dimensions of mesonic operators, see in particular \cite{Erler:2005nr, Mann:2006rh, Correa:2008av, Correa:2009dm} for our system in the ${\cal N} =2$ supersymmetric case $\theta_1 = \theta_2 =0$.
  It remains to be seen whether the susy-breaking rotation preserves integrability.

Another direction for future work is to study the actual tachyon condensation process 
on the field theory side. In the bulk, after tachyon condensation the intersecting $D7$ branes recombine (see {\it e.g.}  \cite{Epple:2003xt}).
For small $\lambda$, the tachyon vacuum corresponds on the field theory side to the local minimum of the one-loop effective potential.
It would be interesting to expand the Lagrangian around the minimum and relate this field theory calculation
to the bulk phenomenon of brane recombination.

\section*{Acknowledgements}

It is pleasure to thank Igor Klebanov, William Linch III, Andrei Parnachev, Martin Rocek and Warren Siegel for  useful discussions. 
This work is supported in part by the DOE grant DEFG-0292-ER40697 and by the NSF grant PHY-0653351-001. 
Any opinions, findings, and conclusions or recommendations expressed in this material are those of the authors
and do not necessarily reflect the views of the National Science Foundation.

\label{global}

\appendix

\section{The Supersymmetric Field Theory \label{appA} }

In this appendix we spell out our conventions and write the ${\cal N} = 2$ supersymmetric action for the usual $D3/D7$ system \cite{Karch:2002sh}.
We first present the Lagrangian in $\NN=1$ superspace  and then in components.  
 Unless otherwise stated, we follow the superspace notations of \cite{Gates:1983nr}.

As familiar, the ${\cal N}=4$ vector multiplet decomposes into an ${\cal N} = 1$ vector multiplet,
\be
V = \bar \theta \sigma^\mu  \theta
A\,_{\mu} + i \theta^2 \bar \theta \bar \lambda - i \bar{ \theta}^2 \theta \lambda  + \theta^2 \bar{ \theta}^2 D \, \quad \mbox{(Wess-Zumino gauge) }
\ee
and  three chiral multiplets
\be
 \Phi^{a} = \phi^a + \theta
\chi^a - \theta^2 F^{a} \, , \quad a = 1,2,3 \, ,
\ee
all in the adjoint representation of the $SU(N)$ gauge group. For zero theta angle, the superspace Lagrangian reads 
\be
\label{Lag4}
{\cal L}_{\NN=4} = \Tr \left[  \int d^4\theta\, e^{-gV}\, \bar{\Phi}_a \,e^{gV} \,\Phi^a +
 \int  d^2\theta\, W^2 + \left( \frac{i\,g}{3!} \int  d^2\theta\, \epsilon_{abc}\,\Phi^a \left[ \Phi^b,\,\Phi^c  \right]  + h.c.       \right) \right] \, ,
\ee
where   $W_\alpha \equiv i \bar{D}^2 D_\alpha V  $ is the usual field strength chiral superfield.
 In this $\NN = 1$ language, only an $SU(3)_R \times U(1)_r$ subgroup of the
 $SU(4)_R$ $R$-symmetry is visible. The $SU(3)_R$ rotates the three chiral superfields leaving $V$ invariant,
 while the $U(1)_r$ is the usual ${\cal N} = 1$ $R$-symmetry, with the chiral superfields having charge $2/3$. 
 \footnote{Note that we are making a graphical distinction between this $U(1)_r$ symmetry and the $U(1)_R$ symmetry  defined in (\ref{choice}).  See the footnote in Appendix B.}

 In components\footnote{In going from superspace to components, we redefine the coupling, $g_{superspace} = \sqrt{2} g_{components}$, to recover the usual normalization.},
\begin{displaymath}
{\cal L}_{\mathcal{N} = 4} = \Tr \Bigg[ -\frac{1}{4}F^{\mu\nu}F_{\mu\nu} - i\bar{\lambda}_{A}\bar{\sigma}^{\mu}D_{\mu}\lambda^{A} - \frac{1}{2}D^{\mu}\bar{X}_{AB} D_{\mu}X^{AB} \Bigg.
 \end{displaymath}
 \be
\Bigg. +i\,\sqrt{2}\, g\,X^{AB}\, \bar{\lambda}_{A}\bar{\lambda}_{B}  -i\,\sqrt{2}\,  g\,\bar{X}_{AB} \lambda^{A}\lambda^{B} - \frac{g^2}{4}[X^{AB}, X^{CD}]\,[\bar{X}_{CD}, \bar{X}_{AB}] \Bigg] \,, 
\ee
where $A, B = 1, \dots,4$. The scalars $X^{AB}$  are related to the three complex scalars $\phi^a$ as
\be
X^{A B} = \left(\begin{array}{cc|cc}
0 & \phi^3 & \phi^2 & \phi^1
\\
-\phi^3 & 0 & \phi^{*}_1 & -\phi^{*}_2
\\
\hline
-\phi^2 & -\phi^{*}_1 & 0 & \phi^{*}_3
\\
-\phi^1 & \phi^{*}_2 &-\phi^{*}_3 & 0
\end{array}\right)
\ee
and obey the self-duality constraint (\ref{selfduality}).

We can also think the ${\cal N} = 4$ vector multiplet as an ${\cal N} = 2$ vector multiplet
(comprising $V$ and $\Phi^3$) and an ${\cal N} = 2$ hyper multiplet (comprising $\Phi^1$ and $\Phi^2$).
We wish to couple  the ${\cal N} = 2$ vector multiplet $(V, \Phi^3)$ to $N_f$ ``flavor'' hyper multiplets in the fundamental
representation of the gauge group. 
Each $\NN=2$ flavor hyper multiplet decomposes into
two ${\cal N} = 1$
chiral multiplets 
\bea && Q = q +  \theta
\psi - \theta^2 f     \quad    
\mbox{and}    \quad    \tilde{Q} =\tilde{ q} +    \theta
\tilde{\psi} - \theta^2 \tilde{f}\, , \nonumber \eea
where $Q$ is in the fundamental representation of $SU(N)$ and $\tilde Q$ in the antifundamental representation. 
In ${\cal N} =1$ superspace, the flavor part of the Lagrangian
reads\footnote{Strictly speaking, this is the Lagrangian for gauge group $U(N)$. For $SU(N)$
 there is a $O(1/N)$ correction  to the $Q^4$ potential, which we neglect since we are interested in the large $N$ limit.}
\be
\label{hyp}
{\cal L}_{hyper} =    \int  d^4\theta \, \bar{Q}_i \,e^{gV}\, Q^i  + \int  d^4\theta\,  \tilde{Q}_i\, e^{-gV}\, \bar{\tilde{Q}}^i  + \left( g  \int d^2\theta \, \tilde{Q}_i \, \Phi^3 \,Q^i   + h.c.       \right) \, ,
\ee
where $i=1,\dots,N_f$ is a flavor index. In components,
\begin{displaymath}
{\cal L}_{hyper} =  -D^{\mu}\bar{Q}_{\II\,i}D_{\mu}Q^{\II\,i} - i\bar{\psi}_i\bar{\sigma}^{\mu}D_{\mu}\psi^i - i\tilde{\psi}_i\sigma^{\mu}D_{\mu}\bar{\tilde{\psi}}\,^i
\end{displaymath}
\begin{displaymath}
\left. -\sqrt{2}\,i\,g\,\tilde{\psi}_i X^{12} \psi^i + \sqrt{2}\,i\,g\,\bar{\psi}_i\bar{X}_{12}\bar{\tilde{\psi}}\,^i \right.
\end{displaymath}
\begin{displaymath}
\left. + ig\sqrt{2}\bar{Q}_{\II\,i}\bar{\Lambda}^{\II}\bar{\tilde{\psi}}\,^i - ig\sqrt{2}\tilde{\psi}_i\Lambda_{\II}Q^{\II\,i} + ig\sqrt{2}\bar{Q}_{\II\,i}\epsilon^{\II \JJ}\Lambda_{\JJ}\psi^i - ig\sqrt{2}\bar{\psi}_i\bar{\Lambda}^{\II}\epsilon_{\II \JJ}Q^{\JJ\,i}\right.
\end{displaymath}
\begin{displaymath}
 - \frac{1}{2}g^2\bar{Q}_{\II\,i}\bar{X}_{AB}X^{AB}Q^{\II\,i} - g^2\bar{Q}_{\JJ\,i}\mathcal{X}_{\II \hat\KK}\mathcal{X}^{ \JJ \hat\KK}Q^{\II\,i} \,  .
\end{displaymath}
\be
 - \frac{g^2}{2}(\bar{Q}_{\II\,i} \cdot Q^{\JJ\,j})(\bar{Q}_{\JJ\,j} \cdot Q^{\II\,i}) - g^2\epsilon_{\II \KK}\epsilon^{\LL \JJ}(\bar{Q}_{\LL\,i} \cdot Q^{\KK\,j})(\bar{Q}_{\JJ\,j} \cdot Q^{\II\,i})\,.
\ee
Following \cite{Erler:2005nr}, we have introduced the $SU(2)_R$ doublets
\be
 Q^{\II} \equiv \left(\begin{array}{c} q
\\
\tilde{q}^*
\end{array}\right)  \, , \quad \Lambda_{\II} \equiv \left(\begin{array}{c} \lambda_1
\\
\lambda_2
\end{array}\right) = \left(\begin{array}{c} \lambda
\\
-  \chi_3
\end{array}\right) \, , \quad \II = 1,2 \,.
\ee
The other  two Weyl spinors can be assembled into an $SU(2)_L$ doublet,
\be
\hat{\Lambda}_{\hat \II} \equiv \left(\begin{array}{c} \lambda_3
\\
\lambda_4
\end{array}\right)= \left(\begin{array}{c} -  \chi_2
\\
-  \chi_1
\end{array}\right) \, ,
\ee  
which does not couple to the flavor hyper multiplets.   Note that to avoid cluttering  we keep color indices  implicit.
Color contractions are almost always obvious. When ambiguity may arise,
we indicate the contraction with a dot. For example in the term
\be
(\bar{Q}_{\II\,i} \cdot Q^{\JJ\,j})(\bar{Q}_{\JJ\,j} \cdot Q^{\II\,i}) 
\ee
the first pair is color contracted, and so is the second pair. 

The $Q^4$ term in the Lagrangian can be written more compactly by introducing flavor-contracted
composite operators, in the adjoint of the gauge group,
\be
\MM_{\JJ \, \, \, \gaugeind{b}}^{\, \, \II \gaugeind{a} }\equiv \frac{1}{\sqrt{2}} Q_{\JJ \mbox{ }i}^{\mbox{ }\gaugeind{a} }\,\bar{Q}_{\mbox{ }\gaugeind{b}}^{\II \mbox{ }i} \, ,
\ee
which may be  decomposed  into the $SU(2)_{R}$ singlet and triplet combinations
  \begin{equation} \label{M1M3}
\MM_{{\bf {1}}} \equiv \MM^{\, \,  \II}_{\II}\quad\mbox{and}\quad\MM_{ {\bf {3} } \JJ  }^{\quad \II}  \equiv
\MM^{\, \, \II}_{\JJ}-\frac{1}{2}\MM^{\, \, \KK}_{\KK}\, \delta^{\II}_{\JJ} \, .
\end{equation}
In terms of the component fields $q$ and $\tilde{q}$,
\bea
 \label{diagonal basis}
 && \MM_{\bf{1}}  =  \frac{1}{\sqrt{2}}  \left(  q \,  \bar{q} +  \bar{\tilde{q}} \,  \tilde{q}   \right)    \\ \nonumber 
&& \MM_{\bf{3}^+} = q \, \tilde{q}   \\ \nonumber 
&&\MM_{\bf{3}^0} =   \frac{1}{\sqrt{2}} \left(     q \,  \bar{q} -   \bar{\tilde{q}}  \,  \tilde{q}    \right) \\ \nonumber
&&\MM_{\bf{3}^-} = \tilde{q}^* \, q^*  \, , \nonumber \eea
where the superscripts refer to the eigenvalues under the Cartan generator of $SU(2)_R$.
The $F$ and $\bar F$ auxiliary fields couple to  ${\cal M}_{{\bf 3}^\pm}$, while the $D$ auxiliary
field couples to ${\cal M}_{{\bf 3}^0}$. Thus the $Q^4$ scalar potential is the square of the triplet composite,
\be
{\cal L}_{Q^4} =- g^2 \, {\rm Tr} \, {\cal M}_{\bf 3} \,   {\cal M}_{\bf 3} \equiv
 -g^2  \, {\rm Tr} \left[  2 {\cal M}_{{\bf 3}^+}  {\cal M}_{{\bf 3}^-} + {\cal M}_{{\bf 3}^0}  {\cal M}_{{\bf 3}^0} \right] \, .
\ee
Finally, let us write the $Q^4$ potential using the gauge-invariant mesonic operators. 
The explicit expressions of the mesons in components are
\bea
 && \OO^{ij}_{\bf{1}}  =  \frac{1}{\sqrt{2}}  \left(  q^{i \, \gaugeind{a}} \,  \bar{q}_{j \, \gaugeind{a}} +  \bar{\tilde{q}}^{i \, \gaugeind{a}}  \,  \tilde{q}_{j \, \gaugeind{a}}   \right)    \\ \nonumber 
&& \OO^{ij}_{\bf{3}^+} = q^{i \, \gaugeind{a}}  \, \tilde{q}_{j \, \gaugeind{a}}   \\ \nonumber 
&&\OO^{ij}_{\bf{3}^0} =   \frac{1}{\sqrt{2}} \left(     q^{i \, \gaugeind{a}}  \,  \bar{q}_{j \, \gaugeind{a}} -   \bar{\tilde{q}}^{i \, \gaugeind{a}}   \,  \tilde{q}_{j \, \gaugeind{a}}    \right) \\ \nonumber
&&\OO^{ij}_{\bf{3}^-} =  \tilde{q}^*\,^{i \, \gaugeind{a}}  \, q^*_{j \, \gaugeind{a}} \,.  \nonumber 
\eea
Withe these definitions, 
\be
{\cal L}_{Q^4} =- \frac{g^2}{2} \, {\rm Tr} \left( 3\, {\cal O}_{\bf 1}^{ij} \,   {\cal O}_{\bf 1}^{ij}- {\cal O}_{\bf 3}^{ij} \,   {\cal O}_{\bf 3}^{ij}\right)  \, .
\ee

\section{The Field Theory for General Angles}

In this appendix we derive the Lagrangian dual to the system with two flavor branes at general angles,
up to an ambiguity in the $Q^4$ terms of the scalar potential, for which we give a general parametrization.
 As explained
in the text, we need to rotate the  $\NN=4$  fields (including in principle the auxiliary fields)
 in the terms of the Lagrangian where  they are coupled
to the second  hyper multiplet.

\subsection{R-symmetry rotations of the ${\cal N}=4$ fields}

Rotation of the $X_m$ scalars in the 49 plane (with angle $\theta_1$) and in the 85 plane (with angle $\theta_2$)
is performed by the matrix
\be
\label{Xrot}
{\cal R}^{( {\bf 6} )} (\theta_1, \theta_2) =  \left(\begin{array}{cccccc} 
 \cos{\theta_1 } & 0 &  0\quad  &\,  0\,&0&-\,   \sin{\theta_1 } 
 \\
0&  \cos{\theta_2 } &0 \quad  &\,  0 \,& \sin{\theta_2 }  & 0
\\ 
0& 0 & 1 \quad &\,  0\,&0& 0
\\ 
0& 0 & 0 \quad & \, 1\,&0& 0
\\ 
0&-\, \sin{\theta_2 } & 0\quad & \, 0\, & \cos{\theta_2 } &  0
 \\
\sin{\theta_1 }  &0& 0\quad & \, 0 \,&0&  \cos{\theta_1 }
\end{array}\right) 
\ee 
A short calculation using the the Clebsh-Gordon  coefficients  (\ref{Xmatrix})  
 gives the corresponding $SU(4)_R$ transformation for the fermions $\lambda_A$,
 \footnote{For completeness, we also list the $SU(4)_R$ transformations corresponding to various $U(1)$ subgroups.
The subgroup $U(1)_R$ (see (\ref{choice})) corresponds to an 89 rotation:
\be
\label{89}
 \left(\begin{array}{cc|cc} e^{i\theta_{89}/2} & & &
\\  & e^{i\theta_{89}/2} & & \\
\hline   & &e^{-i\theta_{89}/2}  &
\\  & & & e^{-i\theta_{89}/2}
\end{array}\right) \,.
\ee
Rotations in the 45 and 67 planes are given respectively by
\be
\label{4567}
 \left(\begin{array}{cc|cc} e^{i\theta_{45}/2} & & &
\\  & e^{-i\theta_{45}/2} & & \\
\hline   & &e^{-i\theta_{45}/2}  &
\\  & & & e^{i\theta_{45}/2}
\end{array}\right)  \quad  \mbox{and} \quad   \left(\begin{array}{cc|cc} e^{i\theta_{67}/2} & & &
\\  & e^{-i\theta_{67}/2} & & \\
\hline   & &e^{i\theta_{67}/2}  &
\\  & & & e^{-i\theta_{67}/2}
\end{array}\right)\,.
\ee
Finally the $U(1)_r$ symmetry of $\NN=1$ superspace is
\be
r =  \left(\begin{array}{c|ccc}
e^{-ir} & & &
 \\
 \hline
 & e^{+ir/3} &  & 
 \\
  & & e^{+ir/3}  & 
\\
 & & & e^{+ir/3} 
\end{array}\right)  \,.
\ee
 }
\be
\label{fermrot}
{\cal R}^{( {\bf 4} )} (\theta_1, \theta_2) =
   \left(\begin{array}{cccc} 
 \cos{\left(\frac{\theta_1 - \theta_2 }{2}\right)} & 0 & i  \sin{\left(\frac{\theta_1 - \theta_2 }{2}\right)} & 0
 \\
0&   \cos{\left(\frac{\theta_1 + \theta_2 }{2}\right)} & 0 & i  \sin{\left(\frac{\theta_1 + \theta_2 }{2}\right)} 
 \\
i \sin{\left(\frac{\theta_1 - \theta_2 }{2}\right)} & 0 &  \cos{\left(\frac{\theta_1 - \theta_2 }{2}\right)} & 0
 \\
0&  i \sin{\left(\frac{\theta_1 + \theta_2 }{2}\right)} & 0 &  \cos{\left(\frac{\theta_1 + \theta_2 }{2}\right)}
\end{array}\right)  \,.
\ee
Ideally, at this point we would provide  the corresponding $SU(4)_R$ transformation of the $F$ and $D$ auxiliary fields.
An unsuccessful  attempt to find such transformation rules using the formalism of \cite{Marcus:1983wb, Gates:1983nr} is described in Appendix C.

Clearly, the $\NN=4$ part does not depend on the angles, since $SU(4)_R$ is an exact symmetry,
\be
{\cal L}_{total} (\theta)   =   {\cal L}_{ {\cal N} = 4} + {\cal L}_{hyper} (\theta)  \,.
\ee
We write
\be
\label{rotatedL}
 {\cal L}_{hyper} (\theta)   =     {\cal L}_{kin}   +  {\cal L}^{(1)}_{Yukawa}  + \LL^{(1)}_{\bar Q X^2 Q}  + {\cal L}^{(2)}_{Yukawa} (\theta) + \LL^{(2)}_{\bar Q X^2 Q} (\theta) 
 + {\cal L}_{Q^4} (\theta)  \, , 
 \ee
 where the superscripts $(1)$ and $(2)$ refer to the first and second flavor, respectively. We have
 indicated which terms are $\theta$ dependent. The  terms  ${\cal L}^{(2)}_{Yukawa} (\theta)$  and  $\LL^{(2)}_{\bar Q X^2 Q} (\theta)$
are fixed unambiguously by the transformations (\ref{Xrot}, \ref{fermrot}). 
By contrast to determine ${\cal L}_{Q^4} (\theta)$ we would need an off-shell superspace formulation for the ${\cal N}=4$  multiplet, which is not available at present.

 The terms that we {\it can} fix are:
\be
{\cal L}_{kin} =  -D^{\mu}\bar{q}_{i}D_{\mu}q^{i} -D^{\mu}\tilde{q}_{i}D_{\mu}\bar{\tilde{q}}^{i} - i\bar{\psi}_i\bar{\sigma}^{\mu}D_{\mu}\psi^i - i\tilde{\psi}_i\sigma^{\mu}D_{\mu}\bar{\tilde{\psi}}\,^i
\ee

\bea
 {\cal L}^{(1)}_{Yukawa} =& 
- 
\sqrt{2}ig\,\tilde{\psi}_1\,\lambda_{1}\,q^1
-
\sqrt{2}ig\,\tilde{q}_1\,\lambda_1\,\psi^1
-
  \sqrt{2}ig
\,\bar{\psi}_1 \,\bar{\lambda}^{2}\,q^1
-
\sqrt{2}ig \,\tilde{\psi}_1\,\lambda_{2}\,\tilde{q}^{*}\,^1     &
\nonumber \\ 
&   -i\,g\,\tilde{\psi}_1 \left(X_8 + i X_9 \right)\psi^1
\, + \,  h.c.     &
\eea

 \bea
 \LL^{(1)}_{\bar Q X^2 Q}&=& -i\,g^2\, \,q^{*}_1\,\Big( \left[ \,X_4\,,\,X_6 \,\right] + i\left[ \,X_4\,,\,X_7 \,\right] +i\left[ \,X_5\,,\,X_6 \,\right] -\left[ \,X_5\,,\,X_7 \,\right]  \Big)\,\tilde{q}^{*}\,^1
\\
\nonumber &&
-i\,g^2\,
\tilde{q}_1\,\Big( \left[ \,X_4\,,\,X_6 \,\right] - i\left[ \,X_4\,,\,X_7 \,\right] -i\left[ \,X_5\,,\,X_6 \,\right] -\left[ \,X_5\,,\,X_7 \,\right]  \Big)\,q^1
\\
\nonumber && +\,i\,g^2\,q^{*}_1\, \Big( \left[\,X_4\,,\,X_5 \,\right]  +\,\left[\,X_6\,,\,X_7 \,\right]  \Big)\, q^1
\\
\nonumber && -\,i\,g^2\,\tilde{q}_1\, \Big( \left[\,X_4\,,\,X_5 \,\right]  +\,\left[\,X_6\,,\,X_7 \,\right]  \Big)\,\tilde{q}^{*}\,^1
\\
\nonumber 
  &&   -\,g^2\,
q^{*}_1\,\Big(X_8^2 + X_9^2  \Big)\,q^1
 -\,g^2\, \tilde{q}_1 \,\Big(X_8^2 + X_9^2 \Big)\,\tilde{q}^{*}\,^1
 \eea

\bea
 {\cal L}^{(2)}_{Yukawa} (\theta) &=& 
- 
\sqrt{2}ig\,\cos{\left(\frac{\theta_1 - \theta_2 }{2}\right)}\,\tilde{\psi}_2\,\lambda_{1}\,q^2
+ \sqrt{2}g\,\sin{\left(\frac{\theta_1 - \theta_2 }{2}\right)}\,\tilde{\psi}_2\,\lambda_{3}\,q^2
  \\
\nonumber &-&
\sqrt{2}ig\,\tilde{q}_2\,\cos{\left(\frac{\theta_1 - \theta_2 }{2}\right)} \,\lambda_1\,\psi^2 
+\sqrt{2}g\,\tilde{q}_2\,\sin{\left(\frac{\theta_1 - \theta_2 }{2}\right)} \,\lambda_3\,\psi^2
   \\
\nonumber
 &-&
 \sqrt{2}ig\,  \cos{\left(\frac{\theta_1 + \theta_2 }{2}\right)}\,\bar{\psi}_2 \,\bar{\lambda}^{2}\,q^2  
- \sqrt{2}g\,  \sin{\left(\frac{\theta_1 + \theta_2 }{2}\right)}\,\bar{\psi}_2 \,\bar{\lambda}^{4}\,q^2
\\ \nonumber 
 &-&
\sqrt{2}ig\,\cos{\left(\frac{\theta_1 + \theta_2 }{2}\right)}\,\tilde{\psi}_2\,\lambda_{2}\,\tilde{q}^{*}\,^2
+ \sqrt{2}g\, \sin{\left(\frac{\theta_1 + \theta_2 }{2}\right)}\,\tilde{\psi}_2\,\lambda_{4}\,\tilde{q}^{*}\,^2
\\ \nonumber
&-&   i\,g\,\tilde{\psi}_2 \left(\cos{\theta_2 }\,X_8 +  \sin{\theta_2 } X_5 + i \cos{\theta_1 }\,X_9 -  i \sin{\theta_1 } X_4 \right)\psi^2      +   h.c.
\eea

 \bea
 \LL^{(2)}_{\bar Q X^2 Q}(\theta)&=&  -\,i\,g^2\,q^{*}_2\,\Big( \cos{\theta_1}\left( \left[ \,X_4\,,\,X_6 \,\right] + i\left[ \,X_4\,,\,X_7 \,\right] \right)+ \cos{\theta_2}\left( i\left[ \,X_5\,,\,X_6 \,\right] -\left[ \,X_5\,,\,X_7 \,\right] \right) \Big)\,\tilde{q}^{*}\,^2 \nonumber
\\
\nonumber && 
-\,i\,g^2\,
\tilde{q}_2\,\Big( \cos{\theta_1}\left( \left[ \,X_4\,,\,X_6 \,\right] - i\left[ \,X_4\,,\,X_7 \,\right] \right)+ \cos{\theta_2}\left(- i\left[ \,X_5\,,\,X_6 \,\right] -\left[ \,X_5\,,\,X_7 \,\right] \right) \Big)\,q^2 \nonumber
\\
\nonumber 
&&  -\,i\,g^2\,q^{*}_2\,\Big(- \sin{\theta_1}\left( \left[ \,X_9\,,\,X_6 \,\right] + i\left[ \,X_9\,,\,X_7 \,\right] \right)   +  \sin{\theta_2}\left(i\left[ \,X_8\,,\,X_6 \,\right] -\left[ \,X_8\,,\,X_7 \,\right] \right) \Big)\,\tilde{q}^{*}\,^2 \nonumber
\\
\nonumber && 
-\,i\,g^2\,
\tilde{q}_2\,\Big( - \sin{\theta_1}\left( \left[ \,X_9\,,\,X_6 \,\right] - i\left[ \,X_9\,,\,X_7 \,\right] \right) + \sin{\theta_2}\left(- i\left[ \,X_8\,,\,X_6 \,\right] -\left[ \,X_8\,,\,X_7 \,\right] \right) \Big)\,q^2 \nonumber
\\
\nonumber 
&& +\,i\,g^2\,q^{*}_2\, \Big( \cos{\theta_1} \cos{\theta_2}\left[\,X_4\,,\,X_5 \,\right]  +\,\left[\,X_6\,,\,X_7 \,\right] -\sin{\theta_1} \sin{\theta_2}\,\left[\,X_8\,,\,X_9 \,\right]  \Big)\, q^2
\\
\nonumber 
&& +\,i\,g^2\,q^{*}_2\,  \Big( \cos{\theta_1} \sin{\theta_2}\left[\,X_4\,,\,X_8 \,\right]  -\sin{\theta_1} \cos{\theta_2}\,\left[\,X_5\,,\,X_9 \,\right]  \Big) \, q^2
\\
\nonumber 
&& -\,i\,g^2\,\tilde{q}_2\, \Big( \cos{\theta_1} \cos{\theta_2}\left[\,X_4\,,\,X_5 \,\right]  +\,\left[\,X_6\,,\,X_7 \,\right] -\sin{\theta_1} \sin{\theta_2}\,\left[\,X_8\,,\,X_9 \,\right]  \Big)\, \tilde{q}^{*}\,^2
\\
\nonumber 
&& -\,i\,g^2\,\tilde{q}_2\,  \Big( \cos{\theta_1} \sin{\theta_2}\left[\,X_4\,,\,X_8 \,\right]  -\sin{\theta_1} \cos{\theta_2}\,\left[\,X_5\,,\,X_9 \,\right]  \Big) \, \tilde{q}^{*}\,^2
\\
&&
  -\,g^2\, q^{*}_2 \left(\,\Big|  \cos{\theta_2 }\,X_8 +  \sin{\theta_2 } X_5\Big|^2 +\Big| \cos{\theta_1 }\,X_9 - \sin{\theta_1 } X_4  \Big|^2 \right)\,q^2
\nonumber \\
&&
  -\,g^2\,  \tilde{q}_2 \left(  \,\Big|  \cos{\theta_2 }\,X_8 +  \sin{\theta_2 } X_5\Big|^2 +\Big|\left(  \cos{\theta_1 }\,X_9 - \sin{\theta_1 } X_4 \right) \Big|^2 \right)\,\tilde{q}^{*}\,^2   \, \, .\nonumber 
 \eea

 \subsection{Parametrizing the $Q^4$ potential}

Writing
\be
 -{\cal L}_{Q^4} (\theta_1, \theta_2) = V_{Q_1^4} + V_{Q_2^4}+ V_{Q_1^2 Q_2^2}(\theta_1,\theta_2) \, ,
\ee
it is clear that
the terms  $V_{Q_1^4}$ and  $V_{Q_2^4}$ (involving respectively only the scalars of the  first and  second hyper multiplet) 
are  unaffected by the rotation. Indeed if we set to zero one of the two hyper multiplets we must recover the standard supersymmetric Lagrangian
(possibly after a change of variables:  if we set to zero the {\it first} hyper multiplet in (\ref{rotatedL}) we must rotate back the $X$ and $\lambda$ fields
to restore $\bar Q X^2 Q$ and Yukawa terms to the standard form).

Recall (Appendix A) that the potential for a single flavor $Q_1$  is
\be
V_{Q_1^4} ={\rm Tr}\, \left[2\, \MM^{11}_{{\bf 3}^+} \MM^{11}_{{\bf 3}^-}+ \MM^{11}_{{\bf 3}^0} \MM^{11}_{{\bf 3}^0}\right] \, ,
\ee
where ${\cal M}_{\bf 3}^{11} \sim (Q_{i=1}\bar Q^{i=1})_{\bf 3}$ is the color--adjoint composite in the triplet of $SU(2)_R$  containing only scalars in the first flavor $i=1$ (compare with (\ref{M1M3})),
and similarly of course for  $V_{Q_2^4}$, with  ${\cal M}_{\bf 3}^{11} \to {\cal M}_{\bf 3}^{22}\sim (Q_{i=2}\bar Q^{i=2})_{\bf 3}$.
The mixed terms can be parametrized by two unknown functions of the angles,
\be \label{Vmixed}
V_{Q_1^2 Q_2^2} = 
2\,g^2 \,{\rm Tr} \,\Big[   f\left( \theta_1 , \theta_2 \right) \left( \MM^{11}_{{\bf 3}^+} \MM^{22}_{{\bf 3}^-}+\MM^{11}_{{\bf 3}^-}\MM^{22}_{{\bf 3}^+}\right)   
  + d\left( \theta_1 , \theta_2 \right) \, \MM^{11}_{{\bf 3}^0} \MM^{22}_{{\bf 3}^0}  \Big]  \,. 
\ee
We have imposed neutrality of the potential under a Cartan generator $U(1) \subset SU(2)_R$. 
This $U(1)$ is preserved for general $\theta_1$, $\theta_2$ and corresponds geometrically to rotations in the 67 plane.
When one of the two angles is zero, say $\theta_2=0$, an $SU(2)$ symmetry is preserved, corresponding geometrically
to rotations in the 567 directions, which is a certain  diagonal combination of $SU(2)_L$ and $SU(2)_R$. The hyper multiplets are neutral under $SU(2)_L$, so under a 567 rotation they just  undergo just an $SU(2)_R$. It follows
that for $\theta_2=0$ the $Q^4$ must be $SU(2)_R$ invariant  
and the functions $f$ and $d$ are related as
\be
\label{fd}
 f\left( \theta , 0 \right)=d\left( \theta , 0\right) \, . 
\ee
The only assumption we have made in writing (\ref{Vmixed}) is that the rotation does not introduce any terms containing
the $SU(2)_R$ singlet composites ${\cal M}_{\bf 1}^{11}$ and ${\cal M}_{\bf 1}^{22}$. This is generally 
the case if the rotated Lagrangian can be obtained from {\it some} off-shell superspace formulation of ${\cal N}=4$ SYM
with manifest $SU(4)_R$ symmetry.  Indeed we know that for zero angles the ${\cal N}=4$ auxiliary fields  only
couple to  ${\cal M}_{\bf 3}$, and rotating the auxiliary fields can never generate ${\cal M}_{\bf 1}$.

When $\theta_1 =\theta_2= \theta$ $\NN=1$ supersymmetry is preserved, and
the $SU(4)$ $R$-symmetry transformation corresponds to a matrix ${\cal R}^{({\bf 4})} \subset SU(3)$,
 which acts on  $F^a$ leaving $D$ invariant. This is a manifest symmetry of the $\NN=1$ superspace formulation.
As reviewed in Appendix C, in this special case one can unambiguously find 
\bea
\label{N=1}
& f\left( \theta , \theta \right) = \cos{\theta }    &
\\ \nonumber
& d\left( \theta , \theta \right) = 1   \, .    &
\eea

Further  constraints follow from  discrete symmetries.
The 89 reflection
$X_8,  X_9   \rightarrow  -X_8 , -X_9$ corresponds
\be
\theta_1 \,, \theta_2  \rightarrow  - \theta_1 \,  , -\theta_2\, .
\ee
Invariance of the $Q^4$ potential under this parity symmetry implies
\bea
f\left( \theta_1 , \theta_2 \right) = f\left( - \theta_1 , - \theta_2 \right)
\\
d\left( \theta_1 , \theta_2 \right) = d\left( - \theta_1 , - \theta_2 \right)\,.
 \eea
 Similarly, invariance under the discrete symmetry $X_4 \leftrightarrow X_5$ and $X_8  \leftrightarrow X_9$, or
\be
\theta_1  \leftrightarrow \theta_2  \,  ,
\ee
implies
\be
\label{12sym}
f\left( \theta_1 , \theta_2 \right) = f\left( \theta_2 , \theta_1 \right) \quad \mbox{and}  \quad  d\left( \theta_1 , \theta_2 \right) = d\left( \theta_2 , \theta_1 \right)   \, .
\ee
Unfortunately this set of relations is not sufficient to fix the functions $f$ and $d$ uniquely.

A crucial assumption we shall make is {\it positivity} of the classical $Q^4$ potential. It would follow from the existence
of a superspace formulation with manifest $SU(4)_R$ symmetry:
 the scalar potential would be proportional to 
 the square of some auxiliary fields,  and it would thus be positive even the susy-breaking $SU(4)_R$ rotation in the terms that couple  the auxiliary fields 
 to the second hyper multiplet. Assuming positivity, we have
\be
\label{positivity}
V_{Q^4} \geq 0  \quad   \Rightarrow  \quad   f\left( \theta_1 , \theta_2 \right)\leq 1 \quad \mbox{and} \quad     d\left( \theta_1 , \theta_2 \right) \leq 1  \quad    \forall  \,  \theta_1, \theta_2\,.
\ee

For small angles, taking into account the discrete symmetries, we can expand
\bea
f\left( \theta_1 , \theta_2 \right)& =& 1-\alpha (\theta_1^2+\theta_2^2) -\beta \theta_1 \theta_2+ O(\theta^3)
\\
d\left( \theta_1 , \theta_2 \right)&=& 1- \tilde \alpha \left( \theta_1^2+ \theta_2^2 \right) - \tilde{\beta}\theta_1 \theta_2 +O(\theta^3)
\eea
for some coefficients $\alpha, \beta, \tilde \alpha,  \tilde{\beta}$.
Imposing (\ref{N=1}) gives $2\alpha+\beta = \frac{1}{2}$ and   $2\tilde \alpha+\tilde \beta =0$, while
 (\ref{fd}) gives $\alpha = \tilde \alpha$. This leaves us with with a single unknown coefficient,
\bea
f\left( \theta_1 , \theta_2 \right)& =&1-\alpha (\theta_1-\theta_2)^2 -\frac{1}{2}  \theta_1 \theta_2+ O(\theta^3)
\\
d\left( \theta_1 , \theta_2 \right)&=& 1- \alpha \left( \theta_1- \theta_2 \right)^2  +O(\theta^3)\, .
\eea  
Positivity of the $Q^4$ potential implies that $\alpha >0$.

Finally we record the explicit expressions of the $Q^4$ terms,  both in terms of component fields,\footnote{We use dots as shorthand notation for color contractions.  Using $\gaugeind{a}, \gaugeind{b} $ for the color indices,
and suppressing all other indices, we set 
\be 
\left( q\,q^* \right) \cdot \left(  q\,q^*  \right) \equiv
\left( q^\gaugeind{a}\,q^*_\gaugeind{b} \right) \cdot \left(  q^\gaugeind{b}\,q^*_\gaugeind{a}  \right) \,,\quad \left( q\,q^* \right)^2 \equiv
\left( q^\gaugeind{a}\,q^*_\gaugeind{b} \right) \cdot \left(  q^\gaugeind{b}\,q^*_\gaugeind{a}  \right)  \,.
\ee
 }
\bea
\label{the potential}
  {\cal L}_{Q^4} (\theta_1, \theta_2)   &=& -2\,g^2\left(\tilde{q}_1\cdot \tilde{q}^*\,^1\right)\left( q^*_1\cdot q^1\right) -2\,g^2\left(\tilde{q}_2\cdot \tilde{q}^*\,^2\right)\left(q^*_2 \cdot q^2 \right)
 \\
&&  -2\,g^2 f\left( \theta_1 , \theta_2 \right)\,\Big[  \left(\tilde{q}_1\cdot \tilde{q}^*\,^2\right)\left(q^*_2\cdot q^1\right) +\left(\tilde{q}_2\cdot \tilde{q}^*\,^1\right)\left(q^*_1\cdot q^2 \right) \Big]
\nonumber \\
&&
-\frac{g^2}{2}\,\left( q^1\,q^*_1-\tilde{q}^*\,^1\,\tilde{q}_1 \right)^2 -\frac{g^2}{2} \left(  q^2\,q^*_2 -\tilde{q}^*\,^2\,\tilde{q}_2 \right)^2
\nonumber \\
&&
  -   g^2 d\left( \theta_1 , \theta_2 \right)  \left( q^1\,q^*_1-\tilde{q}^*\,^1\,\tilde{q}_1 \right) \cdot \left(  q^2\,q^*_2 -\tilde{q}^*\,^2\,\tilde{q}_2 \right)\,,
\nonumber 
\eea
and in terms of the gauge invariant mesonic operators,
\bea
{\cal L}_{Q^4} (\theta_1, \theta_2)& =& 
- \frac{g^2}{2} \, {\rm Tr} \Big[   3\, {\cal O}_{\bf 1}^{11} \,   {\cal O}_{\bf 1}^{11} -{\cal O}_{\bf 3}^{11}  \,   {\cal O}_{\bf 3}^{11} \Big]
- \frac{g^2}{2} \, {\rm Tr} \Big[   3\, {\cal O}_{\bf 1}^{22} \,   {\cal O}_{\bf 1}^{22} - {\cal O}_{\bf 3}^{22}  \,   {\cal O}_{\bf 3}^{22} \Big]
\nonumber \\
&&  -\,g^2\left(d\left( \theta_1 , \theta_2 \right)+2\, f\left( \theta_1 , \theta_2 \right)\right)\, {\rm Tr} \Big[   {\cal O}_{\bf 1}^{12} \,   {\cal O}_{\bf 1}^{21}   \Big]
\nonumber \\
&&  -\,g^2\left(d\left( \theta_1 , \theta_2 \right)-2\, f\left( \theta_1 , \theta_2 \right) \right)\, {\rm Tr} \Big[   {\cal O}_{{\bf 3}^0}^{12} \,   {\cal O}_{{\bf 3}^0}^{21}   \Big]
\nonumber \\
&&
 +   g^2 d\left( \theta_1 , \theta_2 \right)  {\rm Tr} \Big[   {\cal O}_{{\bf 3}^+}^{12}  \,   {\cal O}_{{\bf 3}^-}^{21}  +   {\cal O}_{{\bf 3}^-}^{12}  \,   {\cal O}_{{\bf 3}^+}^{21}        \Big]
\, .
\eea

\section{R-symmetry in ${\cal N} =1$ Superspace}

In this appendix we describe an attempt to derive the $Q^4$ potential for general
angles, using a formalism developed in \cite{Marcus:1983wb, Gates:1983nr}
to describe the general global transformations of ${\cal N} = 4$ SYM in ${\cal N} =1$
superspace language.  The attempt fails, for reasons that could have been anticipated:
while the formalism prescribes how auxiliary fields must transform under general R-symmetry
transformations so that the action is invariant, the transformations do not close off-shell.
Nevertheless we believe that the exercise contains some relevant lessons in the search
of a more complete superspace formulation of ${\cal N} = 4$ SYM and we
 reproduce it here for the benefit
of the technically inclined reader.  We also present 
an application of the formalism to the analogous problem
for ${\cal N} = 2 $ SYM coupled to ${\cal N} =1$ chiral matter: how
to break supersymmetry by inequivalent embeddings of two ${\cal N} =1$ subalgebras into ${\cal N} = 2$.  
In this  case the formalism works,  because the algebra of global transformations closes off-shell.
The simplified problem is interesting in its own right and provides a model for how things
should work in the yet-to-be-found improved superspace formulation of ${\cal N} = 4$.

In $\NN =1$ superspace, $\NN=2$ SYM has a manifest $U(1)_r \times U(1)_u$ subgroup of the $SU(2)_R \times U(1)_R$ R-symmetry,
and
 $\NN = 4$ SYM a manifest $SU(3)_R \times U(1)_r$ subgroup of the $SU(4)_R$ R-symmetry.
Nevertheless, the remaining R-symmetry transformations,
while realized non-linearly, are legitimate off-shell symmetries of the superspace action. They close
off-shell for $\NN=2$ but not for $\NN = 4$.
The explicit transformations
rules were originally given in \cite{Marcus:1983wb}. We follow the presentation of \cite{Gates:1983nr}. Here
we review the superspace formalism of \cite{Marcus:1983wb, Gates:1983nr}, translate it into components and apply it to our problem.

 \subsection{Global symmetries in $\NN = 1$ superspace}

$\NN=1$ supersymmetric theories are invariant under translations, supersymmetry transformations and (under certain conditions) 
R-symmetry transformations. The parameters of these transformations can  be assembled
into a single $x$-independent real superfield $\zeta$,
subject to the gauge-invariance
\be
\delta \zeta = i (\bar \xi - \xi) \, ,
\ee
where $\xi$ is an $x$-independent chiral superfield. The physical components of $\zeta$ (in Wess-Zumino gauge) are
\be
 \zeta_{\alpha \dot{\alpha}} = \frac{1}{2}\left[  \bar{D}_{\dot{\alpha}} , D_\alpha\right] \zeta \arrowvert \,, \quad
\epsilon_\alpha = i \bar{D}^2D_\alpha \zeta   \arrowvert \, , \quad
r =  \frac{1}{2} D^\alpha \bar{D}^2 D_\alpha\, \zeta | \,.
\ee
The vector $\zeta_{\alpha \dot \alpha}$ parametrizes  the translations, the spinor $\epsilon_\alpha$ 
the supersymmetry transformations and the scalar $r$  the $U(1)_r$ symmetry.

Let us next consider the $\NN = 2$ SYM theory.
In $\NN =1$ superspace, the field content consists of a vector superfield $V$ and a chiral superfield $\Phi$. It was 
shown \cite{Gates:1983nr} that the $\NN = 2$ SYM action is invariant under the global transformations
\bea
\label{superR2}
&  \delta \Phi  = -W^{\alpha} \nabla_{\alpha} \eta
-i \,\bar{\nabla}^2\left(\nabla^\alpha \zeta \right)\nabla_\alpha \Phi    &
\\
& e^{-V} \delta e^{V} =   i \left(\bar{\eta}\, \Phi - \eta\,\tilde{ \Phi} \right) +\left(W^{\alpha}\nabla_{\alpha}+\bar{W}^{\dot{\alpha}} \bar{\nabla}_{\dot{\alpha}} \right)\zeta  & \, .   \nonumber
\eea
Moreover, the algebra of these transformations  closes off-shell.  
The parameters $\eta$ and $\zeta$ are $x$-independent, color neutral superfields.  The symbol
$\nabla_\alpha$ denotes the gauge-covariant derivative, $\nabla_\alpha \equiv  e^{-V} D_\alpha e^V$, and $\tilde \Phi  \equiv e^{-V}\bar{\Phi} e^V$.
Note that $\nabla_\alpha \eta = D_\alpha \eta$ and $\nabla_\alpha \zeta = D_\alpha \zeta$. Another obvious
global symmetry of the action is a phase rotation of $\Phi$,
\be \label{U1u}
\delta \Phi =i u \Phi \,,  \quad \delta V = 0 \,. 
\ee
 The parameter $\zeta$ is the real superfield of $\NN=1$ transformations, as above. Its vectorial
 component $\zeta_{\alpha \dot \alpha}$ parametrizes translations, its
 spinorial component  $\epsilon^1_\alpha$ parametrizes the manifest $\NN = 1$ supersymmetry,
 and its auxiliary component $r$ parametrizes the manifest $U(1)_r$ symmetry 
 \be
 \Phi (y, \theta) \to e^{-2 i r} \Phi(y, e^{ir} \theta) \, , \quad   V (x, \theta, \bar \theta) \to V(x, e^{ir} \theta, e^{-ir} \bar \theta )\,.
 \ee
 The parameter $\eta$ is a chiral superfield  that mixes $V$ and $\Phi$. Its components are
\be
z = \eta \arrowvert \, , \quad  \epsilon^2_\alpha = D_\alpha \eta   \arrowvert \, , \quad \mu=D^2 \eta  |\,.
\ee
The complex scalar $z$ corresponds to the central charge, the spinor $\epsilon^2_\alpha$
parametrizes  the second (non-manifest in $\NN=1$ superspace) supersymmetry transformation, and finally the complex scalar $\mu$
 parametrizes the non-manifest  internal symmetries $U(2)/ (U(1)_r \times U(1)_u)$. The complete internal symmetry group of the classical action is $U(2)$,
 parametrized by $u$, $r$ and $\mu$.
 (Note that together the parameters $\zeta$ and $\eta$ 
 form an $\NN = 2$ vector multiplet, mimicking the field content of the theory.)

From (\ref{superR2}, \ref{U1u}),  we find the following global symmetry transformation rules on the  component
fields and auxiliary fields:
\bea
\delta \phi &  = &   - 2\,i \, r\, \phi  +i\,u\, \phi \\
\delta \chi &  = & - \mu \lambda  - i \, r\,  \chi   +   i\,u\, \chi \\
\delta \lambda  & = & \bar{\mu}  \chi  - i \,r\, \lambda  \nonumber \\
 \delta F  & = & 2\,i\, \mu D  +i\,u\, F
\\
 \delta D & = & i \left(  \bar{\mu} F -  \mu \bar{F} \right)  \nonumber \,.
\eea
The two U(1)s in these transformations are  the $r$-symmetry of  $\NN=1$ superspace $U(1)_r$ and the global phase rotation  $U(1)_u$: these
 are natural symmetries from an $\NN=1$ superspace point of view.  The $\NN =2$ SYM theory
   has a $SU(2)_R \times U(1)_R$ R-symmetry  where the $U(1)_R$ and the diagonal $T_3\subset SU(2)_R$ are related to $r$ and $u$ as
   $R=r-\frac{u}{2}$,  $\mu_3= u$, so that we can equivalently write
 \bea
\delta \phi &  = &   -  2 \,i \, R\, \phi 
\\
\delta \chi &  = & - \mu \lambda  -  i \, R\,  \chi   +   i\,\frac{\mu_3}{2}\, \chi \\
\delta \lambda  & = & \bar{\mu}  \chi  -   i\,R\, \lambda-   i\,\frac{\mu_3}{2}\, \lambda  \nonumber \\
\delta F  & = & 2\,i\, \mu D  -i\,\mu_3\,F   \label{N2Rtr}
\\
 \delta D & = & i \left(  \bar{\mu} F -  \mu \bar{F} \right)  \nonumber \,.
\eea
Note that the auxiliary fields transform as a triplet under $SU(2)_R$.

We finally come to  $\NN = 4$ SYM. The $\NN = 4$ action
is invariant under the global transformations \cite{Gates:1983nr}
\bea
\label{superR}
\delta \Phi^a & =& -\left( W^{\alpha} \nabla_{\alpha} \eta^a+ \epsilon^{abc}\bar{\nabla}^2\bar{\eta}_b \tilde{\Phi}_c
  \right) 
 \\ \nonumber
&&
-i\left[ \bar{\nabla}^2\left(\nabla^\alpha \zeta \right)\nabla_\alpha \Phi^a  + \frac{2}{3} \bar{\nabla}^2\left(\nabla^2 \zeta \right) \Phi^a \right]
\eea
\be
e^{-V} \delta e^{V} =   i \left(  \bar{\eta}_a \Phi^a  - \eta^a \tilde{\Phi}_a    \right) + \left(W^\alpha \nabla_\alpha +\bar{W}^{\dot{\alpha}} \bar{\nabla}_{\dot{\alpha}} \right)\zeta \,.
\ee
Unlike the $\NN = 2$ case, the algebra does not close off-shell.
The parameters $\zeta$ and $\eta^a$, $a=1,2,3$  have the same interpretation as before. The real superfield
$\zeta$ contains the parameters of the manifest symmetries, while the chiral superfields $\eta^a$ contain the parameters
of the non-manifest symmetries. In particular their auxiliary auxiliary components of $\mu^a$ are the parameters
of the $SU(4)/ (SU(3) \times U(1)_r)$ R-symmetries.
From (\ref{superR}), after some algebra we find the following $SU(4)/SU(3) $  transformation rules on the component
fields,
\bea
\label{Xcoset}
\delta \phi^a &  = & - \epsilon^{abc} \bar{\mu}_b \bar{\phi}_c - i \frac{2}{3} r\, \phi^a \\
\label{fermicoset}
\delta \chi^a &  = & - \mu^a \lambda  +\frac{1}{3} i \, r\,  \chi^a \\
\delta \lambda  & = & \bar{\mu}_a  \chi^a  - i \,r\, \lambda  \nonumber \\
\label{auxcoset}
 \delta F^a  & = & 2\,i\, \mu^a D  + i  \frac{4}{3}r\, F^a
\\
 \delta D & = & i \left(  \bar{\mu}_a F^a -  \mu^a \bar{F}_a\right)  \nonumber \,.
\eea

\subsection{Application to  $\NN=2$}

Let us consider
$\NN=2$ SYM coupled to $N_f$ $\NN=1$ chiral multiplets $Q^i$, $i =1, \dots N_f$.
The only term in the $\NN =1$ superspace Lagrangian that couples the different flavors  is the K\"ahler term for the chiral multiplets,
\be
\int  d^4\theta \, \bar{Q}_i \,e^{gV}\, Q^i \, .
\ee
(There is no superpotential term  that preserves gauge invariance and the $U(N_f)$ global flavor symmetry.)
In component language, and before integrating out the auxiliary fields, the relevant terms in the Lagrangian are
\be
{\cal L} = \dots + D\left( q^*_i q^i + \left[ \phi \, , \, \bar{\phi} \right] \right) + D^2 + \bar{F}F + \dots
\ee
We then perform an off diagonal $SU(2)_R$ transformation (\ref{N2Rtr}) with $\mu =i\,{\theta/2}$  to the $\NN=2$ SYM component fields that couple to the second chiral multiplet.
\be
{\cal L} = \dots + D\left( q^*_1 q^1+ q^*_2 q^2 \cos{\theta} + \left[ \phi \, , \, \bar{\phi} \right] \right) + \mbox{Re}(F)q^*_2 q^2 \sin{\theta}+ D^2 + \bar{F}F + \dots
\ee
Integrating out the auxiliary fields, we find  the scalar potential
\be
V_{q^4} = |q_1|^4 +  |q_2|^4 +2\,\cos{\theta} \,|q^*_1 q^2|^2  \, ,
\ee
which is positive definite for any $\theta$ since it is proportional to
$D^2 + \bar{F}F$.

\subsection{Application to $\NN=4$}

From (\ref{Xrot}), we see that for infinitesimal $\theta_1$, $\theta_2$,
\bea
\delta \phi_1 & =&    i\, \left( \frac{\theta_1+\theta_{2}}{2}\right) \phi_3 -  i\, \left(\frac{\theta_1-\theta_{2}}{2}\right) \bar \phi_3
 \\
\delta \phi_ 3& = &  i\, \left( \frac{\theta_1+\theta_{2}}{2}\right) \phi_1 +  i\, \left(\frac{\theta_1-\theta_{2}}{2}\right) \bar \phi_1 \nonumber \\
\delta \phi_2 & = &  0\, . \nonumber 
\eea
The holomorphic part of the variation is an infinitesimal $SU(3)$ rotation $\delta \phi^a = i (\theta_1 +\theta_2) (\hat T_6\, \phi)^a$
generated by the  Lie algebra element
\be
\hat{T}_6=\frac{1}{2} \,\hat{ \lambda}_6 = \frac{1}{2} \, \left(\begin{array}{ccc}
0 &0 & 1
\\
0  &0 & 0
 \\
 1 & 0 &0
\end{array}\right)  \,.
\ee
Comparison with  (\ref{Xcoset}) shows that the antiholomorphic part of the variation is an $SU(4)/SU(3)$ transformation
with parameters
\be
r=\mu_1 = \mu_3 = 0\,,\quad  \mu_2 =\, \pm \, i\, \frac{\theta_2-\theta_{1}}{2} \,.
\ee
Recalling that $F^a$ transform in the ${\bf 3}$ of $SU(3)$, and using  (\ref{auxcoset}) for the transformation
rules under $SU(4)/SU(3)$, we find the corresponding  variations of the auxiliary fields,
\bea
&& \delta F^1 =  i\, \left( \frac{\theta_1+\theta_{2}}{2}\right) F^3
\\
&& \delta F^3 =  i\, \left( \frac{\theta_1+\theta_{2}}{2}\right) F^1 \nonumber
\eea
\bea
&& \delta D = -\left(\theta_1 - \theta_2  \right) \mbox{Re}\left( F_2 \right)
\\ \nonumber 
&& \delta   \mbox{Re}\left( F_2 \right) =  \left(\theta_1 - \theta_2  \right) D\,.
\eea
Their naive exponentiation gives
\be
\label{Dlaw}
 \left(\begin{array}{c}
D
 \\
 \mbox{Re}\left( F_2 \right)
\end{array}\right)_{rot} =     \left(\begin{array}{cc}
\cos{ \left( \theta_1-\theta_{2}\right)} & -\sin{ \left( \theta_1-\theta_{2}\right)}
 \\
\sin{ \left( \theta_1-\theta_{2}\right)} & \cos{ \left( \theta_1-\theta_{2}\right)}
\end{array}\right) \left(\begin{array}{c}
D
 \\
 \mbox{Re}\left( F_2 \right)
\end{array}\right)
\ee
\be
\label{Flaw}
 \left(\begin{array}{c}
F_1
 \\
F_3
\end{array}\right)_{rot} =    \left(\begin{array}{cc}
\cos{ \left( \frac{\theta_1+\theta_{2}}{2}\right)} & i\,\sin{ \left( \frac{\theta_1+\theta_{2}}{2}\right)}
 \\
i\,\sin{ \left( \frac{\theta_1+\theta_{2}}{2}\right)} & \cos{ \left( \frac{\theta_1+\theta_{2}}{2}\right)}
\end{array}\right) 
\left(\begin{array}{c}
F_1
 \\
F_3\end{array}\right)\,.
\ee
Given these explicit transformations for the auxiliary fields we can proceed to derive  the form of the $Q^4$ potential after rotation.
The prescription is to transform the auxiliary fields that couple to the second hyper multiplet, leaving untouched
the auxiliary fields that couple to the first hyper multiplet. This method predicts 
\bea
\label{N1df}
&& f\left( \theta_1 , \theta_2 \right) = \cos\left( \frac{\theta_1 + \theta_2}{2} \right)   
\\ \nonumber
&& d\left( \theta_1 , \theta_2 \right) = \cos\left( \theta_1-  \theta_2 \right) \,
\eea
for the parameter functions introduced in (\ref{Vmixed}). This result is clearly incorrect. It does not satisfy
condition (\ref{fd}). Moreover $f(\theta, 0)$ has the wrong periodicity -- the potential should come back to itself
after a $2\pi$ rotation. Since the transformation rules do not close off-shell, it was not permissible
to simply exponentiate the infinitesimal variations.  It appears that this is a fundamental flaw of  this formalism
and that what is required is a different superspace formulation where the $SU(4)_R$ closes off-shell.

\section{Anomalous Dimensions  \label{appB}  }

In this appendix we describe the computation of the one-loop anomalous dimensions of the mesonic operators $\OO^\II\,_\JJ$.
Following \cite{Minahan:2002ve}, we view a single-trace composite operator as a closed spin chain whose sites 
correspond to the elementary fields. In the large $N$ limit and at the one-loop level only nearest neighbor interactions are present. 
The nearest neighbor interaction is conveniently expressed in terms of three elementary operators acting on the vector space of two successive sites. These three operators represent the three independent ways to map two $SU(2)_R$ symmetry indices  of an ``incoming" operator $\OO^\II\,_\JJ$ to the indices of an ``outgoing" operator $\bar{\OO}_\LL\,^\KK$.
They are the trace operator $\mathbb{K}$, the permutation operator $\mathbb{P}$ and the identity operator $\mathbb{I}$:
\be
\mathbb{K}^{ \JJ \LL }_{  \II \KK} \equiv \delta^{\JJ}_{\II}\delta^{\LL}_{\KK}  \, , \qquad \mathbb{P}^{ \JJ \LL }_{  \II \KK} \equiv \delta_{  \II \KK}\delta^{ \JJ \LL }\,,
\qquad \mathbb{I}^{ \JJ \LL }_{  \II \KK} \equiv \delta^{\LL}_{\II}\delta^{\JJ}_{\KK} \,.
\ee

\begin{figure}[t] 
  \centering 
\mbox{\subfigure[]{\epsfig{figure=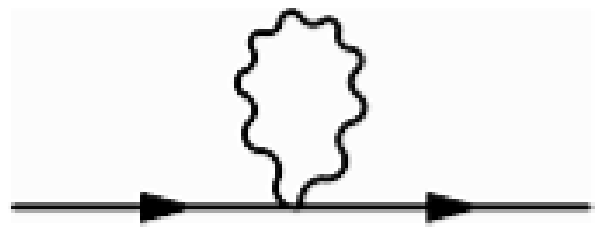,width=4cm,angle=0}}\quad   
  \subfigure[]{\epsfig{figure=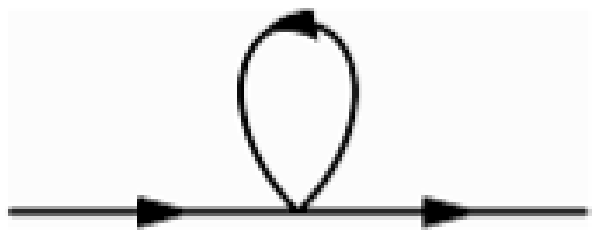,width=4cm,angle=0}}
  \quad   
  \subfigure[]{\epsfig{figure=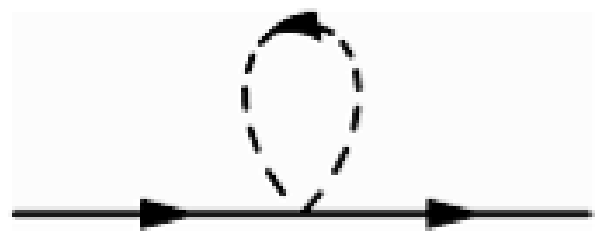,width=4cm,angle=0}}
}
\mbox{
\subfigure[]{\epsfig{figure=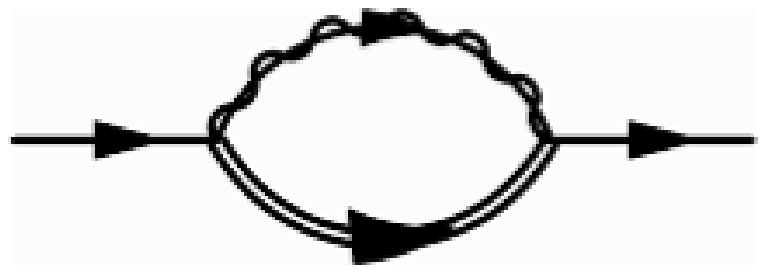,width=4cm,angle=0}} \quad   
  \subfigure[]{\epsfig{figure=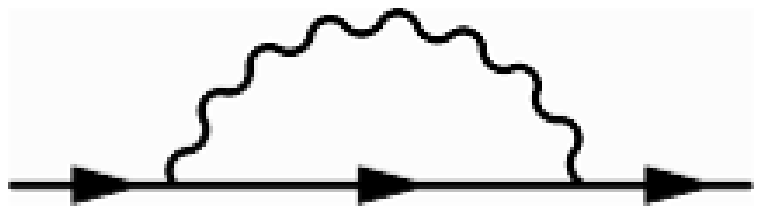,width=4cm,angle=0}}
   } 
   \caption{\it One-loop Feynman diagrams contributing to $\gamma_Q$.} 
  \label{gamma} 
\end{figure}

The anomalous dimension of the mesonic operators 
receives contributions from the Feynman diagrams shown schematically in Figure \ref{renormoper}.
  Since the gauge boson exchange  is $R$-symmetry blind,
  the Feynman diagram shown in figure \ref{renormoper}(a) is proportional to the identity operator,
   \bea
Z_{A}-1 = \left(1-\xi  \right) \frac{g^2 \, N}{8\pi^2} \,\mathbb{I} \,\ln\Lambda  \, \, .
   \eea 
Here $\xi$ is the gauge fixing parameter in the propagator of the gauge boson, which is $\frac{g_{\mu \nu} - \xi \frac{k_{\mu}k_{\nu}}{k^2}}{k^2}$
in our conventions. The $SU(2)_R$ structure of the quartic interaction (figure \ref{renormoper}(b)) is more interesting.
The  scalar vertex has index structure
  \be
\frac{1}{2}\,\delta^{\LL}_{\II}\delta^{\JJ}_{\KK}   -\, \delta^{\JJ}_{\II}\delta^{\LL}_{\KK} = \frac{1}{2}\, \mathbb{I} - \mathbb{K}  \, \, ,
  \ee
 with trace part arising from F-terms  and the identity from the D-terms.
The contribution of  
the quartic interaction to the renormalization of the mesonic operators is then 
     \bea
Z_{Q^4}-1 =  \frac{g^2 \, N}{8\pi^2}\,\left( \mathbb{I}  -\,2\, \mathbb{K} \right)  \, \ln\Lambda    \, \, .
   \eea
Finally, we need to consider the squark self-energy corrections (figure \ref{renormoper}(c), shown in more detail in figure \ref{gamma}).
The contribution of the squark self-energy to the meson renormalization is
\be
Z_{Q}-1 = -\left( 2-\xi \right)\frac{\lambda}{8\pi^2}\,\mathbb{I}\,\ln\Lambda \,.
\ee
We also record for future use the anomalous dimension of the squark,
\be
\label{gammaQ}
\gamma_Q =  \frac{\lambda}{8\pi^2}\,\delta^\JJ_\II\,  \left( 2-\xi \right) \, .
\ee
\smallskip
\smallskip
Adding the diagrams, we find
      \be\label{Zfactor}
   Z
   =1+
   \frac{\lambda}{4\pi^2}\,\mathbb{K} \,\ln\Lambda\,   \ee
and we read off the matrix of anomalous dimensions,
   \be\label{spin}
   \Gamma^{(1)} \equiv \frac{d Z}{d \ln\Lambda} Z^{-1} = \frac{\lambda}{4\pi^2} \,\mathbb{K}  \,.   \ee   
 This answer  is due entirely to the F-terms, since all other contributions (D-terms, gluon exchange 
 and self-energy diagram) add up to zero.
  This is an example of a general property of theories with extended supersymmetry \cite{D'Hoker:1998tz, Constable:2002hw, D'Hoker:2001bq}. 
  
  The trace operator
  is  a $4 \times 4$ matrix that acts on the four dimensional $\bf{2} \times \bar{\bf{2}}$ vector space $Q^\II \bar{Q}_{J}$.  Its eigenstates
  are the  singlet  and the triplet states of $SU(2)_R$,
with eigenvalues $2$ and $0$ respectively. In this basis the anomalous dimension matrix  $ \Gamma^{(1)}$ is diagonal, with eigenvalues:
\be
 \gamma_{\bf{1}} = \frac{\lambda}{2\pi^2}
 \quad \mbox{and}  \quad
 \gamma_{\bf{3}} = 0
\ee
This result is expected  because $\OO_{\bf{3}}$ is an $\NN=2$ chiral primary that obeys the shortening condition $\Delta = 2R$, 
while $\OO_{\bf{1}}$ belongs to a long multiplet and is not protected.

\section{Coleman-Weinberg Potential \label{appD} }

The calculation of the one-loop effective potential 
 is straightforward but somewhat lengthy. 
Here we provide some intermediate steps for the sake of the reader who would like to reproduce our result.
Following the original paper by Coleman and Weinberg \cite{Coleman:1973jx}, the bosonic
and fermionic contributions to the one-loop effective potential are
\be
\label{bosCW}
\VV_{bose} = \frac{1}{64\pi^2} \Tr \, \MM_{b} ^4 \ln{\left(\MM_{b} ^2\right)}
\ee
\be
\label{ferCW}
\VV_{fermi} = -\frac{1}{64\pi^2} \Tr \left(\MM_f \MM_f^\dagger\right)^2 \ln{\left(\MM_f \MM_f^\dagger\right)}    \,   \,  ,
\ee
where the mass matrices  read off by expanding the Lagrangian (\ref{rotatedL}) around the classical background.
We choose the  background (\ref{back})
\be
Q_1 =   \left(\begin{array}{c} q
\\
0
\end{array}\right)   
\, ,\quad
Q_2 =   \left(\begin{array}{c} 0
\\
-q
\end{array}\right) 
\, ,\quad
 \quad q\in \mathbb{C}\,.
\ee
Setting to zero the extra ``double-trace'' couplings, $f \equiv 0$, we find
  the following partial contributions (we write $V \equiv v \, \frac{N\,g^4\,|q|^4}{16\pi^2}\,\ln{|q|^2}$): 
\bea
 && v_Q =  \left(\frac{5+ d^2(\theta_1 , \theta_2)}{2} \right) \, \,  ,
\\  \nonumber
&& v_{fermi} \left( \theta \right) =\left(-\,8
+ 2\,\sin^2{\left(\frac{\theta_1 + \theta_2}{2} \right)} 
+2\,\sin^2{\left(\frac{\theta_1 - \theta_2}{2} \right)} \right) \,  ,
\\ \nonumber
&&  v_A = 3\,  \,  ,
\\ \nonumber
&&  v_X  =
  \left(2-2\,\sin^2{\left(\frac{\theta_1 - \theta_2}{2} \right)}-2\,\sin^2{\left(\frac{\theta_1 + \theta_2}{2} \right)} +4\sin^2{\left(\frac{\theta_1 - \theta_2}{2} \right)}\, \sin^2{\left(\frac{\theta_1 + \theta_2}{2} \right)}
  \right)\,.
\eea
The contribution of the gauge fields is calculated in the Landau gauge\footnote{This is convenient for the following reason. The formula (\ref{bosCW}) arises from the resummation of  polygonal one-loop diagrams with the background fields at the external legs which have zero momenta. When gauge fields are inside the loop there are more diagrams than just the gauge polygons (the polygons are made purely out of gauge fields).  But in the Landau gauge the only diagrams that are non-zero are gauge polygons. Then their contribution is simply (\ref{bosCW}) multiplied by $3$. The extra factor of 3 stems from the trace of the numerator of the Landau gauge propagator.}. 
Adding these partial contributions,
\be
\VV_{1-loop}\left( \theta_1, \theta_2 ; f=0\right)  \equiv 
\frac{\lambda^2}{N} \,v_{1-loop} \,  |q|^4 \ln |q| \,
\ee
with
\be
v_{1-loop} = \frac{1}{8 \pi^2} \left[4\,\sin^2{\left(\frac{\theta_1 + \theta_2}{2} \right)}  \sin^2{\left(\frac{\theta_1 - \theta_2}{2} \right)}   + \frac{ d^2(\theta_1 , \theta_2)-1}{2}   \right]     \, .
\ee
For $\theta_1 = \,\theta_2=\theta$, using $d(\theta, \theta)=1$ we find
\be
\VV_{1-loop}\left( \theta, \theta; f=0 \right) = 0  \, ,
\ee
as expected since this configuration preserves $\NN=1$ susy. 

Along this classical background, the tree-level potential ${\cal V}_{tree} \equiv - ({\cal L}_{Q^4} + \delta {\cal L}_{fund})$ (see (\ref{the potential}) and (\ref{deformation})) evaluates to
\be \label{Vtree}
{\cal V}_{tree} (q) = \frac{ \lambda}{N} \,|q|^4\Big(1- d(\theta_1 , \theta_2)\Big) + \frac{f_{{\bf 3}^+}  }{N}  \, |q|^4    \equiv  \frac{\lambda}{N} \, {\mathcal C}_{\lambda} +  \frac{f_{{\bf 3}^+}}{N}  \, {\mathcal C}_{f} \,.
\ee
The classical background was chosen precisely to ensure that the only double-trace coupling contributing at tree level is $f_{{\bf 3}^+}$.
The Callan-Symanzik equation for the effective potential reads
\be \label{RG}
\left[  \mu \frac{\partial}{\partial \mu} + \beta_f \frac{\partial}{\partial f} + \beta_\lambda \frac{\partial}{\partial \lambda}  - \gamma_q^{(1)} \, q \frac{\partial}{\partial q} \right]  {\cal V}(q, \mu, f, \lambda) = 0 \, .
\ee
We can drop the $\beta_\lambda$ term since it is subleading for large $N$.
Writing ${\cal V}(f, \lambda)= {\cal V}_{tree} + {\cal V}_{1-loop} (f=0) + O(f) + O(\lambda^3)$, the CZ equation 
  allows to extract the $f$-independent one-loop coefficient of $\beta_f$, $\beta_f (f=0) = a(\lambda)= a^{(1)} \lambda^2  + O(\lambda^3)$,
 \bea
 a^{(1)}  &=&  
  4\, \gamma_q^{(1)}\left( {\mathcal C}_{\lambda}/{\mathcal C}_{f}\right) + v_{1-loop} 
\\  \nonumber 
  &=&
   \frac{1}{16\pi^2} \left[  \Big(1-d(\theta_1 , \theta_2)  \Big) 
 + \frac{1}{2} \Big(1-d(\theta_1 , \theta_2)  \Big)^2
+ 4\,\sin^2{\left(\frac{\theta_1 + \theta_2}{2} \right)}  \sin^2{\left(\frac{\theta_1 - \theta_2}{2} \right)} \right] \,.
 \eea
  To study the stability of the CW potential, we must first perform the standard RG improvement \cite{Coleman:1973jx}.
A detailed discussion for the problem at hand can be found in section 3 of \cite{Pomoni:2008de}. One finds that the ``perturbative vacuum'' $q=0$   is stable
if and only if  $\beta_f$ admits real zeros. In our case $\beta_f$ has {\it imaginary} zeros and symmetry breaking does occur.
This is one of the manifestations of the tachyonic instability on the field theory side.

\bibliographystyle{JHEP}
\bibliography{D3D7bib}

\providecommand{\href}[2]{#2}\begingroup\raggedright\begin{thebibliography}{10}

\bibitem{Sen:2004nf}
A.~Sen, {\it {Tachyon dynamics in open string theory}},  {\em Int. J. Mod.
  Phys.} {\bf A20} (2005) 5513--5656,
  [\href{http://xxx.lanl.gov/abs/hep-th/0410103}{{\tt hep-th/0410103}}].

\bibitem{Karch:2002sh}
A.~Karch and E.~Katz, {\it {Adding flavor to AdS/CFT}},  {\em JHEP} {\bf 06}
  (2002) 043, [\href{http://xxx.lanl.gov/abs/hep-th/0205236}{{\tt
  hep-th/0205236}}].

\bibitem{Sakai:2003wu}
T.~Sakai and J.~Sonnenschein, {\it {Probing flavored mesons of confining gauge
  theories by supergravity}},  {\em JHEP} {\bf 09} (2003) 047,
  [\href{http://xxx.lanl.gov/abs/hep-th/0305049}{{\tt hep-th/0305049}}].

\bibitem{Kuperstein:2004hy}
S.~Kuperstein, {\it {Meson spectroscopy from holomorphic probes on the warped
  deformed conifold}},  {\em JHEP} {\bf 03} (2005) 014,
  [\href{http://xxx.lanl.gov/abs/hep-th/0411097}{{\tt hep-th/0411097}}].

\bibitem{Arean:2004mm}
D.~Arean, D.~E. Crooks, and A.~V. Ramallo, {\it {Supersymmetric probes on the
  conifold}},  {\em JHEP} {\bf 11} (2004) 035,
  [\href{http://xxx.lanl.gov/abs/hep-th/0408210}{{\tt hep-th/0408210}}].

\bibitem{Ouyang:2003df}
P.~Ouyang, {\it {Holomorphic D7-branes and flavored N = 1 gauge theories}},
  {\em Nucl. Phys.} {\bf B699} (2004) 207--225,
  [\href{http://xxx.lanl.gov/abs/hep-th/0311084}{{\tt hep-th/0311084}}].

\bibitem{Canoura:2006es}
F.~Canoura, J.~D. Edelstein, and A.~V. Ramallo, {\it {D-brane probes on
  L(a,b,c) superconformal field theories}},  {\em JHEP} {\bf 09} (2006) 038,
  [\href{http://xxx.lanl.gov/abs/hep-th/0605260}{{\tt hep-th/0605260}}].

\bibitem{Apreda:2006bu}
R.~Apreda, J.~Erdmenger, D.~Lust, and C.~Sieg, {\it {Adding flavour to the
  Polchinski-Strassler background}},  {\em JHEP} {\bf 01} (2007) 079,
  [\href{http://xxx.lanl.gov/abs/hep-th/0610276}{{\tt hep-th/0610276}}].

\bibitem{Sieg:2007by}
C.~Sieg, {\it {Holographic flavour in the N = 1 Polchinski-Strassler
  background}},  {\em JHEP} {\bf 08} (2007) 031,
  [\href{http://xxx.lanl.gov/abs/0704.3544}{{\tt 0704.3544}}].

\bibitem{Penati:2007vj}
S.~Penati, M.~Pirrone, and C.~Ratti, {\it {Mesons in marginally deformed
  AdS/CFT}},  {\em JHEP} {\bf 04} (2008) 037,
  [\href{http://xxx.lanl.gov/abs/0710.4292}{{\tt 0710.4292}}].

\bibitem{Wang:2003yc}
X.-J. Wang and S.~Hu, {\it {Intersecting branes and adding flavors to the
  Maldacena- Nunez background}},  {\em JHEP} {\bf 09} (2003) 017,
  [\href{http://xxx.lanl.gov/abs/hep-th/0307218}{{\tt hep-th/0307218}}].

\bibitem{Nunez:2003cf}
C.~Nunez, A.~Paredes, and A.~V. Ramallo, {\it {Flavoring the gravity dual of N
  = 1 Yang-Mills with probes}},  {\em JHEP} {\bf 12} (2003) 024,
  [\href{http://xxx.lanl.gov/abs/hep-th/0311201}{{\tt hep-th/0311201}}].

\bibitem{Karch:2000gx}
A.~Karch and L.~Randall, {\it {Open and closed string interpretation of SUSY
  CFT's on branes with boundaries}},  {\em JHEP} {\bf 06} (2001) 063,
  [\href{http://xxx.lanl.gov/abs/hep-th/0105132}{{\tt hep-th/0105132}}].

\bibitem{DeWolfe:2001pq}
O.~DeWolfe, D.~Z. Freedman, and H.~Ooguri, {\it {Holography and defect
  conformal field theories}},  {\em Phys. Rev.} {\bf D66} (2002) 025009,
  [\href{http://xxx.lanl.gov/abs/hep-th/0111135}{{\tt hep-th/0111135}}].

\bibitem{Erdmenger:2002ex}
J.~Erdmenger, Z.~Guralnik, and I.~Kirsch, {\it {Four-Dimensional Superconformal
  Theories with Interacting Boundaries or Defects}},  {\em Phys. Rev.} {\bf
  D66} (2002) 025020, [\href{http://xxx.lanl.gov/abs/hep-th/0203020}{{\tt
  hep-th/0203020}}].

\bibitem{Skenderis:2002vf}
K.~Skenderis and M.~Taylor, {\it {Branes in AdS and pp-wave spacetimes}},  {\em
  JHEP} {\bf 06} (2002) 025,
  [\href{http://xxx.lanl.gov/abs/hep-th/0204054}{{\tt hep-th/0204054}}].

\bibitem{Constable:2002xt}
N.~R. Constable, J.~Erdmenger, Z.~Guralnik, and I.~Kirsch, {\it {Intersecting
  D3-branes and holography}},  {\em Phys. Rev.} {\bf D68} (2003) 106007,
  [\href{http://xxx.lanl.gov/abs/hep-th/0211222}{{\tt hep-th/0211222}}].

\bibitem{Adams:2001jb}
A.~Adams and E.~Silverstein, {\it {Closed string tachyons, AdS/CFT, and large N
  QCD}},  {\em Phys. Rev.} {\bf D64} (2001) 086001,
  [\href{http://xxx.lanl.gov/abs/hep-th/0103220}{{\tt hep-th/0103220}}].

\bibitem{Dymarsky:2005nc}
A.~Dymarsky, I.~R. Klebanov, and R.~Roiban, {\it {Perturbative gauge theory and
  closed string tachyons}},  {\em JHEP} {\bf 11} (2005) 038,
  [\href{http://xxx.lanl.gov/abs/hep-th/0509132}{{\tt hep-th/0509132}}].

\bibitem{Dymarsky:2005uh}
A.~Dymarsky, I.~R. Klebanov, and R.~Roiban, {\it {Perturbative search for fixed
  lines in large N gauge theories}},  {\em JHEP} {\bf 08} (2005) 011,
  [\href{http://xxx.lanl.gov/abs/hep-th/0505099}{{\tt hep-th/0505099}}].

\bibitem{Pomoni:2008de}
E.~Pomoni and L.~Rastelli, {\it {Large N Field Theory and AdS Tachyons}},
  \href{http://xxx.lanl.gov/abs/0805.2261}{{\tt 0805.2261}}.

\bibitem{Breitenlohner:1982bm}
P.~Breitenlohner and D.~Z. Freedman, {\it {Positive Energy in anti-De Sitter
  Backgrounds and Gauged Extended Supergravity}},  {\em Phys. Lett.} {\bf B115}
  (1982) 197.

\bibitem{Marcus:1983wb}
N.~Marcus, A.~Sagnotti, and W.~Siegel, {\it {TEN-DIMENSIONAL SUPERSYMMETRIC
  YANG-MILLS THEORY IN TERMS OF FOUR-DIMENSIONAL SUPERFIELDS}},  {\em Nucl.
  Phys.} {\bf B224} (1983) 159.

\bibitem{Gates:1983nr}
S.~J. Gates, M.~T. Grisaru, M.~Rocek, and W.~Siegel, {\it {Superspace, or one
  thousand and one lessons in supersymmetry}},  {\em Front. Phys.} {\bf 58}
  (1983) 1--548, [\href{http://xxx.lanl.gov/abs/hep-th/0108200}{{\tt
  hep-th/0108200}}].

\bibitem{Aharony:1998xz}
O.~Aharony, A.~Fayyazuddin, and J.~M. Maldacena, {\it {The large N limit of N =
  2,1 field theories from three- branes in F-theory}},  {\em JHEP} {\bf 07}
  (1998) 013, [\href{http://xxx.lanl.gov/abs/hep-th/9806159}{{\tt
  hep-th/9806159}}].

\bibitem{Kruczenski:2003be}
M.~Kruczenski, D.~Mateos, R.~C. Myers, and D.~J. Winters, {\it {Meson
  spectroscopy in AdS/CFT with flavour}},  {\em JHEP} {\bf 07} (2003) 049,
  [\href{http://xxx.lanl.gov/abs/hep-th/0304032}{{\tt hep-th/0304032}}].

\bibitem{Epple:2003xt}
F.~T.~J. Epple and D.~Lust, {\it {Tachyon condensation for intersecting branes
  at small and large angles}},  {\em Fortsch. Phys.} {\bf 52} (2004) 367--387,
  [\href{http://xxx.lanl.gov/abs/hep-th/0311182}{{\tt hep-th/0311182}}].

\bibitem{Berkovits:1993zz}
N.~Berkovits, {\it {A Ten-dimensional superYang-Mills action with off-shell
  supersymmetry}},  {\em Phys. Lett.} {\bf B318} (1993) 104--106,
  [\href{http://xxx.lanl.gov/abs/hep-th/9308128}{{\tt hep-th/9308128}}].

\bibitem{Baulieu:2007ew}
L.~Baulieu, N.~J. Berkovits, G.~Bossard, and A.~Martin, {\it {Ten-dimensional
  super-Yang-Mills with nine off-shell supersymmetries}},  {\em Phys. Lett.}
  {\bf B658} (2008) 249--254, [\href{http://xxx.lanl.gov/abs/0705.2002}{{\tt
  0705.2002}}].

\bibitem{Dixon:1986qv}
L.~J. Dixon, D.~Friedan, E.~J. Martinec, and S.~H. Shenker, {\it {The Conformal
  Field Theory of Orbifolds}},  {\em Nucl. Phys.} {\bf B282} (1987) 13--73.

\bibitem{Cvetic:2003ch}
M.~Cvetic and I.~Papadimitriou, {\it {Conformal field theory couplings for
  intersecting D-branes on orientifolds}},  {\em Phys. Rev.} {\bf D68} (2003)
  046001, [\href{http://xxx.lanl.gov/abs/hep-th/0303083}{{\tt
  hep-th/0303083}}].

\bibitem{Abel:2003vv}
S.~A. Abel and A.~W. Owen, {\it {Interactions in intersecting brane models}},
  {\em Nucl. Phys.} {\bf B663} (2003) 197--214,
  [\href{http://xxx.lanl.gov/abs/hep-th/0303124}{{\tt hep-th/0303124}}].

\bibitem{Antoniadis:2000jv}
I.~Antoniadis, K.~Benakli, and A.~Laugier, {\it {Contact interactions in
  D-brane models}},  {\em JHEP} {\bf 05} (2001) 044,
  [\href{http://xxx.lanl.gov/abs/hep-th/0011281}{{\tt hep-th/0011281}}].

\bibitem{Erler:2005nr}
T.~Erler and N.~Mann, {\it {Integrable open spin chains and the doubling trick
  in N = 2 SYM with fundamental matter}},  {\em JHEP} {\bf 01} (2006) 131,
  [\href{http://xxx.lanl.gov/abs/hep-th/0508064}{{\tt hep-th/0508064}}].

\bibitem{Mann:2006rh}
N.~Mann and S.~E. Vazquez, {\it {Classical open string integrability}},  {\em
  JHEP} {\bf 04} (2007) 065,
  [\href{http://xxx.lanl.gov/abs/hep-th/0612038}{{\tt hep-th/0612038}}].

\bibitem{Correa:2008av}
D.~H. Correa and C.~A.~S. Young, {\it {Reflecting magnons from D7 and D5
  branes}},  {\em J. Phys.} {\bf A41} (2008) 455401,
  [\href{http://xxx.lanl.gov/abs/0808.0452}{{\tt 0808.0452}}].

\bibitem{Correa:2009dm}
D.~H. Correa and C.~A.~S. Young, {\it {Asymptotic Bethe equations for open
  boundaries in planar AdS/CFT}},
  \href{http://xxx.lanl.gov/abs/0912.0627}{{\tt 0912.0627}}.

\bibitem{Minahan:2002ve}
J.~A. Minahan and K.~Zarembo, {\it {The Bethe-ansatz for N = 4 super
  Yang-Mills}},  {\em JHEP} {\bf 03} (2003) 013,
  [\href{http://xxx.lanl.gov/abs/hep-th/0212208}{{\tt hep-th/0212208}}].

\bibitem{D'Hoker:1998tz}
E.~D'Hoker, D.~Z. Freedman, and W.~Skiba, {\it {Field theory tests for
  correlators in the AdS/CFT correspondence}},  {\em Phys. Rev.} {\bf D59}
  (1999) 045008, [\href{http://xxx.lanl.gov/abs/hep-th/9807098}{{\tt
  hep-th/9807098}}].

\bibitem{Constable:2002hw}
N.~R. Constable {\em et~al.}, {\it {PP-wave string interactions from
  perturbative Yang-Mills theory}},  {\em JHEP} {\bf 07} (2002) 017,
  [\href{http://xxx.lanl.gov/abs/hep-th/0205089}{{\tt hep-th/0205089}}].

\bibitem{D'Hoker:2001bq}
E.~D'Hoker and A.~V. Ryzhov, {\it {Three-point functions of quarter BPS
  operators in N = 4 SYM}},  {\em JHEP} {\bf 02} (2002) 047,
  [\href{http://xxx.lanl.gov/abs/hep-th/0109065}{{\tt hep-th/0109065}}].

\bibitem{Coleman:1973jx}
S.~R. Coleman and E.~Weinberg, {\it {Radiative Corrections as the Origin of
  Spontaneous Symmetry Breaking}},  {\em Phys. Rev.} {\bf D7} (1973)
  1888--1910.

\end{thebibliography}\endgroup

\end{document}